\documentstyle[epsfig,12pt]{article}
\newcommand{\be}{\begin{eqnarray}}
\newcommand{\ee}{\end{eqnarray}}

\def\lsim{\mathrel{\rlap{\lower4pt\hbox{\hskip1pt$\sim$}}
    \raise1pt\hbox{$<$}}}               
\def\gsim{\mathrel{\rlap{\lower4pt\hbox{\hskip1pt$\sim$}}
    \raise1pt\hbox{$>$}}}               

\begin{document}

\begin{figure}[htb]

\epsfxsize=6cm \epsfig{file=logo_INFN.epsf}

\end{figure}

\vspace{-4.75cm}

\Large{\rightline{Sezione ROMA III}}
\large{
\rightline{Via della Vasca Navale 84}
\rightline{I-00146 Roma, Italy}

\vspace{0.6cm}

\rightline{INFN-RM3 98/2}
\rightline{September 1998}
}

\normalsize{}

\vspace{1.5cm}

\begin{center}

\LARGE{Power corrections in the longitudinal and transverse structure
functions of proton and deuteron}\footnote{\bf To appear in Nuclear
Physics B.}\\

\vspace{1cm}

\large{G. Ricco$^{(a,b)}$, S. Simula$^{(c)}$ and M. Battaglieri$^{(b)}$}\\

\vspace{0.5cm}

\normalsize{$^{(a)}$Dipartimento di Fisica, Universit\'a di Genova\\
Via Dodecanneso 33, I-16146, Genova, Italy\\ $^{(b)}$Istituto Nazionale 
di Fisica Nucleare, Sezione di Genova\\ Via Dodecanneso 33, I-16146, 
Genova, Italy\\$^{(c)}$Istituto Nazionale di Fisica Nucleare, Sezione 
Roma III\\ Via della Vasca Navale 84, I-00146 Roma, Italy}

\end{center}

\vspace{0.5cm}

\begin{abstract}

\noindent Power corrections to the $Q^2$ behaviour of the low-order moments
of both the longitudinal and transverse structure functions of proton and
deuteron have been investigated using available phenomenological fits of
existing data in the $Q^2$ range between $1$ and $20 ~ (GeV/c)^2$. The
Natchmann definition of the moments has been adopted for disentangling
properly target-mass and dynamical higher-twist effects in the data. The
leading twist has been treated at next-to-leading order in the strong
coupling constant and the effects of higher orders of the perturbative
series have been estimated using a renormalon-inspired model. The
contributions of (target-dependent) multiparton correlations to both $ 1 /
Q^2$ and $1 / Q^4$ power terms have been determined in the transverse
channel, while the longitudinal one appears to be consistent with a pure
infrared renormalon picture in the whole $Q^2$-range between $1$ and $20 ~
(GeV/c)^2$. Finally, the extracted twist-2 contribution in the deuteron
turns out to be compatible with the hypothesis of an enhanced $d$-quark
parton distribution at large $x$.

\end{abstract}

\newpage

\pagestyle{plain}

\section{Introduction}

\indent The experimental investigation of deep-inelastic lepton-hadron
scattering has provided a wealth of information on the occurrence of Bjorken
scaling and its violations, giving a decisive support to the rise of the
parton model and to the idea of asymptotic freedom. Quantum Chromodynamics
($QCD$) has been thereby proposed as the theory describing the logarithmic
violations to scaling in the asymptotic region and its predictions at
leading ($LO$) and next-to-leading ($NLO$) orders have been nicely confirmed
by the experiments. However, in the pre-asymptotic region the full
dependence of the hadron structure functions on the squared four-momentum
transfer, $Q^2 \equiv q \cdot q$, is affected also by power corrections,
which can originate from non-perturbative physics and are well beyond the
predictive power of perturbative $QCD$. An important tool for the
theoretical investigation of the $Q^2$ behaviour of the structure functions
is the Operator Product Expansion ($OPE$): the logarithmic scale dependence
is provided by the so-called leading twist operators, which in the parton
language are one-body operators whose matrix elements yield the contribution
of the individual partons to the structure functions. On the contrary, power
corrections are associated to higher-twist operators which measure the
relevance of correlations among partons (see, e.g., \cite{SV82}).

\indent In case of unpolarised inelastic electron scattering the nucleon
response is described by two independent quantities: the transverse
$F_2^N(x, Q^2)$ and the longitudinal $F_L^N(x, Q^2)$ structure functions,
where $x \equiv Q^2 / 2M \nu$ is the Bjorken variable, with $M$ and $\nu$
being the nucleon mass and the energy transfer in the target rest frame.
Systematic measurement of the transverse structure function of the nucleon,
$F_2^N(x, Q^2)$, (more precisely, of the proton and the deuteron
\cite{SLAC,Bosted, EMC,BCDMS,NMC}) have been carried out in the kinematical
range $10^{-4} \lsim x \lsim 1$ and for $Q^2$ values up to several hundreds
of $(GeV/c)^2$; thus, various phenomenological fits of the data sets are
presently available, like those of Refs. \cite{SLAC,Whitlow,Tulay}. As for
the longitudinal to transverse ($L/T$) cross section ratio, $R_{L/T}^N(x,
Q^2) \equiv \sigma_L^N(x, Q^2) / \sigma_T^N(x, Q^2)$, the experimentally
investigated kinematical range is $0.0045 \lsim x \lsim 0.7$ and $1 \lsim
Q^2 ~ (GeV/c)^2 \lsim 70$ \cite{EMC,BCDMS,NMC,RLT}; however, the sparse and
still fluctuating data are not yet sufficient to put significative
constraints on the phenomenological fits (see Refs.
\cite{Whitlow_2,Bartelski}), particularly in the low-$x$ region ($x \lsim
0.3$).

\indent Various analyses of power-suppressed terms in the world data on both
$F_2^N(x, Q^2)$ and $F_L^N(x, Q^2)$ structure functions have been already
carried out; they are based either on the choice of a phenomenological
ans\"atz \cite{Virchaux,Ji} or on renormalon-inspired models
\cite{Webber,Stein}, adopting for the leading twist the $LO$ or $NLO$
approximations. Very recently, also the effects of the
next-to-next-to-leading order ($NNLO$) have been investigated on $F_2^N(x,
Q^2)$ and $R_{L/T}^N(x, Q^2)$ \cite{Bodek} as well as on the
parity-violating structure function $x F_3^N(x, Q^2)$ \cite{Sidorov},
measured in the neutrino and antineutrino scattering off iron by the $CCFR$
collaboration \cite{CCFR}. Present analyses at $NNLO$ seem to indicate that
power corrections in $F_2^N(x, Q^2)$, $R_{L/T}^N(x, Q^2)$ and $x F_3^N(x,
Q^2)$ can be quite small. However, in such analyses the highest value of $x$
is typically limited at $\simeq 0.75$ and also the adopted $Q^2$ range is
quite restricted ($Q^2 \gsim 5 \div 6 ~ (GeV/c)^2$ at $x \simeq 0.75$) in
order to avoid the nucleon-resonance region; therefore, in such kinematical
conditions the power corrections represent always a small fraction of the
leading-twist term (corrected for target-mass effects). Furthermore, when
power corrections are investigated in a narrow $Q^2$-window, it is clear
that variations of the logarithmic dependence of the leading twist can
easily simulate a small power-like term as well as a significative
sensitivity to the higher-twist anomalous dimensions cannot be easily
achieved.

\indent We point out that the smallness of the sum of the various dynamical
higher-twists does not imply the smallness of its individual terms and
therefore the question whether present data are compatible with an
alternate-sign twist expansion is still open. To answer this question, we
extend in this paper the range of values of $Q^2$ down to $Q^2 \simeq 1 ~
(GeV/c)^2$ in order to enhance the sensitivity to power-like terms. For such
low values of $Q^2$ we expect that the effects of target-dependent
higher twists (i.e., multiparton correlations) should show up, thanks also
to the contributions of the nucleon-resonance region and the nucleon elastic
peak. The inclusion of the nucleon-resonance region is clearly worthwhile
also because of parton-hadron duality arguments (see, e.g., Ref.
\cite{RS98}).

\indent In this paper we analyse the power corrections to the $Q^2$
behaviour of the low-order moments of both the longitudinal and transverse
structure functions of proton and deuteron using available phenomenological
fits of existing data in the $Q^2$ range between $1$ and $20 ~ (GeV/c)^2$,
including also the $SLAC$ proton data of Ref. \cite{Bosted} covering the
region beyond $x \simeq 0.7$ up to $x \simeq 0.98$. The Natchmann definition
of the moments is adopted for disentangling properly target-mass and
dynamical higher-twist effects in the data. The leading twist is treated at
$NLO$ in the strong coupling constant and, as far as the transverse channel
is concerned, it is extracted simultaneously with the higher-twist terms
from the data. The effects of higher orders of the perturbative series are
estimated adopting the infrared ($IR$) renormalon model of Ref.
\cite{Webber}, containing both $1 / Q^2$ and $1 / Q^4$ power-like terms. It
turns out that the longitudinal and transverse data cannot be explained
simultaneously by a renormalon contribution only; however, we will show that
in the whole $Q^2$ range between $1$ and $20 ~ (GeV/c)^2$ the longitudinal
channel appears to be consistent with a pure $IR$-renormalon picture,
adopting strengths which are not inconsistent with those expected in the
naive non-abelianization ($NNA$) approximation \cite{Stein} and agree well
with the results of a recent analysis \cite{Sidorov} of the $CCFR$ data
\cite{CCFR} on $x F_3^N(x, Q^2)$ . Then, after including the $1 / Q^2$ and
$1 / Q^4$ $IR$-renormalon terms as fixed by the longitudinal channel, the
twist-4 and twist-6 contributions arising from multiparton correlations are
phenomenologically determined in the transverse channel and found to have
non-negligible strengths with opposite signs. It is also shown that our
determination of multiparton correlation effects in $F_2^N(X, Q^2)$ is only
marginally affected by the specific value adopted for the strong coupling
constant at the $Z$-boson mass, at least in the range between $\simeq 0.113$
and $\simeq 0.118$. Finally, an interesting outcome of our analysis is that
the extracted twist-2 contribution in the deuteron turns out to be
compatible with the enhancement of the $d$-quark parton distribution
recently proposed in Ref. \cite{Bodek}.

\indent The paper is organised as follows. In the next Section the main features of the $OPE$, the $NLO$ approximation of the leading twist and the usefulness of the Natchmann definition of the moments are briefly reminded. In Section 3 our procedure for the evaluation of the {\em experimental} longitudinal and transverse moments is described. Section 4 is devoted to a phenomenological analysis of the data at $NLO$, while the inclusion of the $IR$-renormalon uncertainties is presented in Section 5. The results of our final analysis of the transverse data are collected in Section 6, and our main conclusions are summarised in Section 7.  

\section{The Operator Product Expansion and the Leading Twist}

\indent The complete $Q^2$ evolution of the structure functions can be obtained using the $OPE$ \cite{OPE} of the time-ordered product of the two currents entering the virtual-photon nucleon forward Compton scattering amplitude, viz.
 \be
    T[J(z) ~ J(0)] = \sum_{n,\alpha} ~ f_n^{\alpha}(-z^2) ~ z^{\mu_1} 
    z^{\mu_2} ... z^{\mu_n} ~ O_{\mu_1 \mu_2 ... \mu_n}^{\alpha}
    \label{eq:current}
 \ee
where $O_{\mu_1 \mu_2 ... \mu_n}^{\alpha}$ are symmetric traceless operators of dimension $d_n^{\alpha}$ and twist $\tau_n^{\alpha} \equiv d_n^{\alpha} - n$, with $\alpha$ labelling different operators of spin $n$. In Eq. (\ref{eq:current}) $f_n^{\alpha}(-z^2)$ are coefficient functions, which are calculable in $pQCD$ at short-distance. Since the imaginary part of the forward Compton scattering amplitude is simply the hadronic tensor containing the structure functions measured in deep inelastic scattering ($DIS$) experiments, Eq. (\ref{eq:current}) leads to the well-known twist expansion for the Cornwall-Norton ($CN$) moments of the transverse structure function, viz.
 \be
    \tilde{M}_n^T(Q^2) \equiv \int dx_0^1 ~ x^{n - 2} ~ F_2^N(x, Q^2) = 
    \sum_{\tau = 2, even}^{\infty} E_{n \tau}(\mu, Q^2) ~ O_{n \tau}(\mu) 
    ~ \left( {\mu^2 \over Q^2} \right)^{{\tau - 2 \over 2}}
    \label{eq:CN}
 \ee
where $\mu$ is the renormalization scale, $O_{n \tau}(\mu)$ are the (reduced) matrix elements of operators with definite spin $n$ and twist $\tau$, containing the information about the non-perturbative structure of the target, and $E_{n \tau}(\mu, Q^2)$ are dimensionless coefficient functions, which can be expressed perturbatively as a power series of the running coupling constant $\alpha_s(Q^2)$.

\indent As it is well known, in case of the leading twist ($\tau = 2$) one ends up with a non-singlet quark operator $\hat{O}_{\tau = 2}^{NS}$ with corresponding matrix elements $O_{n 2}^{NS} \equiv a_n^{NS}$ and coefficients $E_{n 2}^{NS}$, and with singlet quark $\hat{O}_{\tau = 2}^S$ and gluon $\hat{O}_{\tau = 2}^G$ operators with corresponding matrix elements $a_n^{\pm}$ and coefficients $E_{n 2}^{\pm}$ (after their mixing under renormalization group equations); explicitly one has
 \be
    \tilde{M}_n^T(Q^2) & = & \mu_n^T(Q^2) + \mbox{higher twists}
    \label{eq:MT_CN} \\
    \mu_n^T(Q^2) & \equiv & \mu_n^{NS}(Q^2) + \mu_n^S(Q^2)
    \label{eq:muT}
 \ee
with at $NLO$
 \be
    \mu_n^{NS}(Q^2) & = & a_n^{NS} ~ \left[ {\alpha_s(Q^2) \over
    \alpha_s(\mu^2)} \right]^{\gamma_n^{NS}} ~ { 1 + \alpha_s(Q^2) 
    R_n^{NS} / 4\pi \over 1 + \alpha_s(\mu^2) R_n^{NS}/ 4\pi}
    \label{eq:muNS_NLO} \\
    \mu_n^S(Q^2) & = & a_n^- ~ \left[ {\alpha_s(Q^2) \over
    \alpha_s(\mu^2)} \right]^{\gamma_n^-} ~ { 1 + \alpha_s(Q^2) R_n^- / 
    4\pi \over 1 + \alpha_s(\mu^2) R_n^- / 4\pi} ~ + \nonumber \\ 
    & & a_n^+ ~ \left[ {\alpha_s(Q^2) \over \alpha_s(\mu^2)} 
    \right]^{\gamma_n^+} ~ { 1 + \alpha_s(Q^2) R_n^+ / 4\pi \over 1 + 
    \alpha_s(\mu^2) R_n^+ / 4\pi}
    \label{eq:muS_NLO}
 \ee
where all the anomalous dimensions $\gamma_n^i$ ($i = NS, \pm$) and coefficients $R_n^i$ can be found in, e.g., Ref. \cite{Altarelli} for $n \leq 10$ and for a number of active flavours, $N_f$, equal to $N_f = 3, 4$ and $5$. At $NLO$ the running coupling constant $\alpha_s(Q^2)$ is given by
 \be
    \alpha_s(Q^2) = {4\pi \over \beta_0 \mbox{ln}(Q^2 /
    \Lambda_{\overline{MS}}^2)} \left \{ 1 - {\beta_1 \over \beta_0^2} 
    {\mbox{ln ln}(Q^2 / \Lambda_{\overline{MS}}^2) \over \mbox{ln}(Q^2 / 
    \Lambda _{\overline{MS}}^2)} \right \}
    \label{alfas}
 \ee
where $\Lambda_{\overline{MS}}$ is the $QCD$ scale in the $\overline{MS}$ scheme, $\beta_0 = 11 - 2N_f/3$ and $\beta_1 = 102 - 38 N_f / 3$. The coefficients $a_n^{\pm}$ can be rewritten as follows
 \be
    a_n^+ & = & (1 - b_n) ~ \mu_n^S(\mu^2) ~ + ~ c_n ~ \mu_n^G(\mu^2) 
    \nonumber \\
    a_n^- & = & b_n ~ \mu_n^S(\mu^2) ~ - ~ c_n ~ \mu_n^G(\mu^2)
    \label{eq:apm}
 \ee
where $\mu_n^G(Q^2)$ is the $CN$ moment of the gluon distribution function
$G(x, Q^2)$ of order $n$, viz. $\mu_n^G(Q^2) = \int_0^1 dx ~ x^{n - 2} ~
G(x, Q^2)$. It turns out that, since quark and gluons are decoupled at large
$x$, one has $b_n \sim 1$ and $c_n \sim 0$ already for $n \gsim 4$, so that
only $a_n^-$ contributes to Eq. (\ref{eq:muS_NLO}). Moreover, again for $n
\gsim 4$ one has $\gamma_n^- \simeq \gamma_n^{NS}$ and $R_n^-(f) \simeq
R_n^{NS}$ \cite{Altarelli}, which implies that for $n \gsim 4$ the evolution
of the leading-twist singlet moments $\mu_n^S(Q^2)$ almost coincide with
that of the non-singlet ones $\mu_n^{NS}(Q^2)$. Therefore, when $n \geq 4$,
we will consider for the leading twist term the following $NLO$ expression
 \be
    \label{eq:muT_NLO}
    \mu_n^T(Q^2) = \mu_n^{NS}(Q^2) + \mu_n^S(Q^2) \to_{n \geq 4} a_n^{(2)} ~ 
    \left[ {\alpha_s(Q^2) \over \alpha_s(\mu^2)} \right]^{\gamma_n^{NS}} ~ 
    {1 + \alpha_s(Q^2) R_n^{NS} / 4\pi \over 1 + \alpha_s(\mu^2) R_n^{NS} / 
    4\pi}
 \ee
with $a_n^{(2)} \equiv \mu_n^{NS}(\mu^2) + \mu_n^S(\mu^2)$. To sum up, the $Q^2$ evolution of the leading-twist transverse moments is completely determined by $pQCD$ for each spin $n$ and all the unknown twist-2 parameters reduce to the three matrix elements $a_2^{NS}$ and $a_2^{\pm}$ in case of the second moment, and only to one matrix element, $a_n^{(2)}$, for $n \geq 4$.

\indent The contribution of power-like corrections to the $Q^2$ dependence of the moments (\ref{eq:CN}) is due to higher twists corresponding to $\tau \geq 4$. Already at $\tau = 4$ the set of basic operators is quite large and their number grows rapidly as $\tau$ increases; moreover, each set of operators mix under renormalization group equations. The short-distance coefficients $E_{n \tau}$ can in principle be determined perturbatively, but the calculations are cumbersome and the large number of the involved matrix elements $O_{n \tau}(\mu)$ makes the resulting expression for the moments of limited use. Therefore, following Ref. \cite{Ji}, in this work we will make use of {\em effective} anomalous dimensions for the higher-twist terms (see next Section).

\indent The $OPE$ can be applied also to the longitudinal structure function $F_L^N(x, Q^2)$, where power corrections are expected to be more important than in the transverse case, because $F_L^N(x, Q^2)$ is vanishing at $LO$. However, $NLO$ effects generate a non-vanishing contribution; more precisely, since the longitudinal structure function is defined as
 \be
    F_L^N(x, Q^2) \equiv F_2^N(x, Q^2) ~ \left(1 + {4 M^2 x^2 \over Q^2} 
    \right) ~ {R_{L/T}^N(x, Q^2) \over 1 + R_{L/T}^N(x, Q^2)}
    \label{eq:FL}
 \ee
where $R_{L/T}^N(x, Q^2) \equiv \sigma_L^N(x, Q^2) / \sigma_T^N(x, Q^2)$ is the $L/T$ cross section ratio, one has
 \be
    \tilde{M}_n^L(Q^2) \equiv \int_0^1 dx ~ x^{n - 2} ~ F_L^N(x, Q^2) = 
    \mu_n^L(Q^2) + \mbox{higher twists}
    \label{eq:ML_CN}
 \ee
with the leading-twist contribution $\mu_n^L(Q^2)$ given at $NLO$ by
 \be
    \mu_n^L(Q^2) = {\alpha_s(Q^2) \over 4 \pi} {1 \over n + 1} \left \{ {8 
    \over 3} \mu_n^q(Q^2) + {2d \over n + 2} \mu_n^G(Q^2) \right \} + 
    \mu_n^c(Q^2)
    \label{eq:muL_NLO}
 \ee
where $\mu_n^q(Q^2)$ is the light-quark flavour contribution to the $n$-th moment of the transverse structure function, $\mu_n^c(Q^2)$ is the charm quark contribution (starting when the invariant produced mass $W$ is greater than twice the mass of the charm quark), and $d = 12/9$ and $20/9$ for $N_f = 3$ and $4$, respectively. In what follows Eq. (\ref{eq:muL_NLO}) will be evaluated using sets of parton distributions available in the literature (see Section 5).

\indent For massless targets only operators with spin $n$ contribute to the $n$-th $CN$ moments of both the longitudinal and transverse structure functions. When the target mass is non-vanishing, operators with different spins can contribute and consequently the higher-twist terms in the expansion of the $CN$ moments $\tilde{M}_n^T(Q^2)$ and $\tilde{M}_n^L(Q^2)$ contain now also target-mass terms, which are of pure kinematical origin. It has been shown by Natchmann \cite{Natchmann} that even when $M \neq 0$ the moments can be redefined in such a way that only spin-$n$ operators contribute in the $n$-th moment, namely
 \be
    M_n^T(Q^2) & \equiv & \int_0^1 dx {\xi^{n+1} \over x^3} F_2^N(x, Q^2) 
    {3 + 3 (n + 1) r + n (n + 2) r^2 \over (n + 2) (n + 3)}
    \label{eq:MT} \\
    M_n^L(Q^2) & \equiv & \int_0^1 dx {\xi^{n+1} \over x^3} \left[ F_L^N(x, 
    Q^2) + {4M^2 x^2 \over Q^2} F_2^N(x, Q^2) {(n + 1) \xi / x - 2(n + 2) 
    \over (n + 2) (n + 3)} \right]
    \label{eq:ML}
 \ee
where $r \equiv \sqrt{1 + 4M^2x^2 / Q^2}$ and $\xi \equiv 2x / (1 + r)$ is the Natchmann variable. Using the {\em experimental} $F_2^N(x, Q^2)$ and $F_L^N(x, Q^2)$ in the r.h.s. of Eqs. (\ref{eq:MT}-\ref{eq:ML}), the target-mass corrections are cancelled out and therefore the twist expansions of the experimental Natchmann moments $M_n^T(Q^2)$ and $M_n^L(Q^2)$ contain only {\em dynamical} twists, viz.
 \be
    M_n^T(Q^2) & = & \mu_n^T(Q^2) + \mbox{dynamical higher twists}
    \label{eq:MT_NAT} \\
    M_n^L(Q^2) & = & \mu_n^L(Q^2) + \mbox{dynamical higher twists}
    \label{eq:ML_NAT}
 \ee
where the leading-twist terms $\mu_n^T(Q^2)$ and $\mu_n^L(Q^2)$ are given at $NLO$ by Eqs. (\ref{eq:muT}-\ref{eq:muS_NLO}) and (\ref{eq:muL_NLO}), respectively.

\section{Data on Transverse and Longitudinal Moments}

\indent For the evaluation of the Natchmann transverse $M_n^T(Q^2)$ (Eq. (\ref{eq:MT})) as well as longitudinal $M_n^L(Q^2)$ (Eq. (\ref{eq:ML})) moments systematic measurements of the experimental structure functions $F_2^N(x, Q^2)$ and $F_L^N(x, Q^2)$ are required in the whole $x$-range at fixed values of $Q^2$. However, such measurements are not always available and therefore we have adopted interpolation formulae ("pseudo-data"), which fit the considerable amount of existing data on the proton and deuteron structure functions.

\indent In case of the transverse channel we have used: ~ i) the Bodek's fit \cite{SLAC,Whitlow} to the inclusive $(e, e')$ $SLAC$ data in the resonance region for values of the invariant produced mass $W$ smaller than $2.5 ~ GeV$, and ~ ii) the Tulay's fit \cite{Tulay} to the world data in the $DIS$ region for $W > 2.5 ~ GeV$. In order to have continuity at $W = 2.5 ~ GeV$ the Bodek's fit has been divided by a factor $1.03$ (at any $Q^2$), which is inside the experimental errors. Moreover, we have also taken into account the wealth of $SLAC$ proton data \cite{Bosted} beyond $x \simeq 0.7$ up to $x \simeq 0.98$ through a simple interpolation fit given in Ref. \cite{Bosted}. All these fits cover the range of $x$ which is crucial for the evaluation of the moments considered in this work; therefore, the uncertainties on the moments are related only to the accuracy of the interpolation formulae; the latter are simply given by a $\pm 4 \%$ total (systematic + statistical) error reported for the $SLAC$ data \cite{SLAC,Bosted} and by the upper and lower bounds of the Tulay's fit quoted explicitly in Ref. \cite{Tulay}. Finally, since the whole set of $DIS$ (unpolarised) $SLAC$ and $BCDMS$ data (on which also the Tulay's fit is based) is known to favour the value $\alpha_s(M_Z^2) \simeq 0.113$ (see Ref. \cite{Virchaux}), in what follows the value $\Lambda_{\overline{MS}} = 290 ~ (240) ~ MeV$ at $N_f = 3 ~ (4)$ will be adopted for the calculation of the running coupling constant at $NLO$ via Eq. (\ref{alfas})\footnote{The impact of a different value of $\alpha_s(M_Z^2)$, much closer to the updated $PDG$ value $\alpha_s(M_Z^2) = 0.119 \pm 0.002$ \cite{PDG98}, will be briefly addressed at the end of Section 5.}.

\indent As for the longitudinal channel, the structure function $F_L^N(x, Q^2)$ is reconstructed via Eq. (\ref{eq:FL}), using for the $L / T$ ratio the phenomenological fit (and the interpolation uncertainties) provided in  Ref. \cite{Bartelski}. In the nucleon resonance region the ratio $R_{L/T}^N(x, Q^2)$ has been taken from Ref. \cite{Burkert} for the lowest values of $Q^2$ (namely $Q^2 = 1 \div 2 ~ (GeV/c)^2$), while it has been assumed to be equal to zero for $Q^2 > 2 ~ (GeV/c)^2$, which is consistent with the limited amount of available data \cite{Stuart}.

\indent Since the $OPE$ is totally inclusive, the contribution of the nucleon elastic channels has to be included in the calculation of the moments, viz.
 \be
    F_2^N(x, Q^2) & = & F_2^{N(inel)}(x, Q^2) + F_2^{N(el)}(x, Q^2)
    \nonumber \\
    F_L^N(x, Q^2) & = & F_L^{N(inel)}(x, Q^2) + F_L^{N(el)}(x, Q^2) 
    \label{eq:Ftotal}
 \ee
and consequently
 \be
    M_n^T(Q^2) & = & \left[ M_n^T(Q^2) \right]_{inel} + \left[ M_n^T(Q^2) 
    \right]_{el} \nonumber \\
    M_n^L(Q^2) & = & \left[ M_n^L(Q^2) \right]_{inel} + \left[ M_n^L(Q^2) 
    \right]_{el} 
    \label{eq:Mtotal}
 \ee
In case of the proton one has
 \be
    \label{eq:F2el}
    F_2^{p(el)}(x, Q^2) & = & \delta(x - 1) ~ {[G_E^p(Q^2)]^2 + \eta ~
    [G_M^p(Q^2)]^2 \over 1 + \eta} \\
    \label{eq:FLel}
    F_L^{p(el)}(x, Q^2) & = & \delta(x - 1) ~ {[G_E^p(Q^2)]^2 \over \eta}
 \ee
where $G_E^p(Q^2)$ and $G_M^p(Q^2)$ are the electric and magnetic (Sachs) proton form factors, respectively, and $\eta \equiv Q^2 / 4 M^2$. Thus, the contribution of the proton elastic channel is explicitly given by
 \be
    \label{eq:MTel}
    \left[ M_n^T(Q^2) \right]_{el}  & \to_{proton} & \left( {2 \over 1 + 
    r^*} \right)^{n + 1} ~ {3 + 3(n + 1) r^* + n(n + 2) r^{*2} \over (n + 2) 
    (n + 3)} \nonumber \\
    & & {[G_E^p(Q^2)]^2 + \eta ~ [G_M^p(Q^2)]^2 \over 1 + \eta} \\
    \label{eq:MLel}
    \left[ M_n^L(Q^2) \right]_{el}  & \to_{proton} & {1 \over 1 + \eta} 
    \left( {2 \over 1 + r^*} \right)^{n + 1} \left \{ [G_E^p(Q^2)]^2 - 
    [G_M^p(Q^2)]^2 + \right. \nonumber \\
    & & \left. {n + 1 \over n + 3} \left [ 1 + {2 \over (n + 2) (1 + r^*)} 
    \right ] {[G_E^p(Q^2)]^2 + \eta ~ [G_M^p(Q^2)]^2 \over 1 + \eta} 
    \right \} 
 \ee
with $r^* = \sqrt{1 + 4 M^2 / Q^2} = \sqrt{1 + 1 / \eta}$.

\indent In case of the deuteron the folding of the nucleon elastic channel with the momentum distribution of the nucleon in the deuteron gives rise to the so-called quasi-elastic peak. In the $Q^2$-range of interest the existing data are only fragmentary and therefore we have computed $F_2^{D(el)}(x, Q^2)$ and $F_L^{D(el)}(x, Q^2)$ using the procedure of Ref. \cite{CS}, which includes the folding of the nucleon elastic peak with the nucleon momentum distribution in the deuteron as well as final state interaction effects; such procedure can be applied both at low ($Q^2 \lsim 1 ~ (GeV/c)^2$) and high values of $Q^2$ (see Ref. \cite{Anghi}). The results of the calculations, performed using the deuteron wave function corresponding to the Paris nucleon-nucleon potential \cite{Paris}, are compared in Fig. 1 with the $SLAC$ data of Ref. \cite{SLAC_deut} at $Q^2 = 1.75$ and $3.25 ~ (GeV/c)^2$. It can be seen that the agreement with the data is quite good for the transverse part and still acceptable for the less accurate longitudinal response.

\indent In Figs. 2 and 3 the $Q^2$ behaviours of the inelastic, elastic and total transverse moments are reported for the proton and the deuteron, respectively, in the whole $Q^2$-range of interest in this work, i.e. $1 \lsim Q^2 (GeV/c)^2 \lsim 20$. Note that in case of the deuteron our data refer to the moments of the structure function per nucleon. It can clearly be seen that the inelastic and elastic contributions exhibit quite different $Q^2$-behaviours and, in particular, the inelastic part turns out to be dominant for $Q^2 \gsim 1 ~ (GeV/c)^2$ in the second moment and only for $Q^2 \gsim n ~ (GeV/c)^2$ in higher order moments. The $Q^2$-behaviour of the longitudinal moments is illustrated in Figs. 4 an 5 in case of the proton and the deuteron, respectively. It can be seen that the elastic contribution drops quite fast and it is relevant only at the lowest values of $Q^2$ ($Q^2 \simeq$ few $(GeV/c)^2$).

\indent We point out that our present determination of the longitudinal and
transverse {\em experimental} moments is clearly limited by the use of
phenomenological fits of existing data (i.e. {\em "pseudo-data"}), which are
required in order to interpolate the structure functions in the whole
$x$-range for fixed values of $Q^2$, as well as by the existing large
uncertainties in the determination of the $L / T$ ratio $R_{L/T}^N(x, Q^2)$.
Thus, both transverse data with better quality at $x \gsim 0.6$ and $Q^2
\lsim 10 ~ (GeV/c)^2$ and more precise, systematic determinations of the $L
/ T$ cross section ratio are still required to improve our experimental
knowledge of the $Q^2$-behaviour of the low-order moments of the nucleon
structure functions.

\section{Analysis of the Transverse Data at $NLO$}

\indent In this Section we present our analysis of the transverse {\em pseudo-data} of Figs. 2 and 3, adopting for the leading twist the $NLO$ expressions (\ref{eq:muNS_NLO}-\ref{eq:muS_NLO}) and for the power corrections a purely phenomenological ans\"atz. Indeed, as already pointed out, several higher-twist operators exist and mix under the renormalization-group equations; such a mixing is rather involved (see, e.g., Ref. \cite{Okawa} in case of twist-4 operators) and in particular the number of mixing operators increases with the spin $n$. Since complete calculations of the higher-twist anomalous dimensions are not yet available, we use the phenomenological ans\"atz already adopted in Ref. \cite{Ji}, which in case of the second moment $M_{n = 2}^T(Q^2)$ leads to the expansion
 \be
    M_2^T(Q^2) = \mu_2^{NS}(Q^2) + \mu_2^S(Q^2) + a_2^{(4)} ~ \left[ 
    {\alpha_s(Q^2) \over \alpha_s(\mu^2)} \right]^{\gamma_2^{(4)}} ~ 
    {\mu^2 \over Q^2}
    \label{eq:M2T}
 \ee
where the logarithmic $pQCD$ evolution of the twist-4 contribution is
accounted for by the term $[\alpha_s(Q^2) /
\alpha_s(\mu^2)]^{\gamma_2^{(4)}}$ with an {\em effective} anomalous
dimension $\gamma_2^{(4)}$ and the parameter $a_2^4$ represents the overall
strength of the twist-4 term at $Q^2 = \mu^2$. For the $QCD$ renormalization
scale $\mu$ we adopt hereafter the value $\mu = 1 ~ GeV/c$. We have
explicitly checked that a different choice for $\mu$ does not modify any
conclusion of this work, because (e.g. in Eq. (\ref{eq:M2T})) the values of
the parameter $a_2^4$ corresponding to two different choices $\mu$ and
$\mu'$ turn out to be related by the logarithmic $pQCD$ evolution, i.e.
$a_2^4(\mu') = a_2^4(\mu) ~ [\alpha_s(\mu'^2) /
\alpha_s(\mu^2)]^{\gamma_2^4} ~ \mu^2 / \mu'^2$.

\indent The unknown parameters appearing in Eq. (\ref{eq:M2T}) (i.e., the
three twist-2 parameters $a_2^{NS} \equiv \mu_2^{NS}(Q^2 = \mu^2)$, $a_2^S
\equiv \mu_2^S(Q^2 = \mu^2)$ and $a_2^G \equiv \mu_2^G(Q^2 = \mu^2)$ and the
two twist-4 parameters $a_2^{(4)}$ and $\gamma_2^{(4)}$) have been
determined by fitting our data for the proton and the deuteron (see Fig.
2(a) and 3(a), respectively), adopting the least-$\chi^2$ procedure in the
$Q^2$-range between $1$ and $20 ~ (GeV/c)^2$. It turned out that the singlet
and gluon twist-2 parameters can be directly determined form the data only
in case of the deuteron, but not in case of the proton. Thus, in fitting the
proton data we have kept fixed the parameters $a_2^S$ and $a_2^G$ at the
values obtained for the deuteron. Our results are shown in Figs. 6(a) and
7(a) in the whole $Q^2$-range of the analysis, while the values obtained for
the various parameters are reported in Table 1, together with the
uncertainties of the fitting procedure corresponding to one-unit increment
of the $\chi^2 / N$ variable (where $N$ is the number of degrees of
freedom). It can clearly be seen that:

\begin{itemize}

\item the twist-4 contribution to second transverse moment $M_2^T(Q^2)$ is
quite small in the proton and is almost vanishing in the deuteron in the
whole $Q^2$-range of our analysis; the latter result suggests that the
twist-4 effect in the neutron comes with a sign opposite to that in the
proton at variance with the expectations from the bag model \cite{JS81};

\item the twist-4 contribution in the proton at $Q^2 = \mu^2 = 1 ~
(GeV/c)^2$ (i.e., $a_2^{(4)} = 0.012 \pm 0.010$) turns out to be
significantly smaller than the result $a_2^{(4)} = 0.030 \pm 0.003$ quoted
in Ref. \cite{Ji}, where however the twist-2 term was not simultaneously
fitted to the data, but instead it was calculated using the parton densities
of Ref. \cite{GRV} evolved at $NLO$. Therefore, we have repeated our
analysis by fixing the twist-2 term at the $GRV$ prediction \cite{GRV}
obtaining $a_2^{(4)} = 0.02 \pm 0.01$, which is still lower but not
inconsistent within the errors with the result of Ref. \cite{Ji}. We point
out that a small variation of the twist-2 term can affect significantly the
strength of the small twist-4 term; that is why our uncertainty from a
simultaneous $a_2^{NS}$ and $a_2^{(4)}$ fit appears to be quite larger than
the one found in Ref. \cite{Ji};

\item as a cross-check, we have also fitted separately the non-singlet
parameter $a_2^{NS}$ to the $"(p - n) / 2"$ data, defined as the difference
between the proton and deuteron data, obtaining $a_2^{NS} = 0.029 \pm
0.009$, $a_2^{(4)} = 0.012 \pm 0.010$ and $\gamma_2^{(4)} = 1 \pm 1$.
Combining these results with those found in the deuteron, we expect to get
in case of the proton $a_2^{NS} = 0.096 \pm 0.017$, $ a_2^{(4)} = 0.012 \pm
0.010$ and $\gamma_2^{(4)} = 1 \pm 1$, which indeed are in nice agreement
with the proton fit results of Table 1.

\end{itemize}

\indent As described in Section 2, in case of the transverse moments $M_n^T(Q^2)$ with $n \geq 4$ the evolution of the twist-2 contribution can be simplified and assumed to be a pure non-singlet one; therefore, we have considered the following twist expansion
 \be
    M_{n \geq 4}^T(Q^2) = \mu_n^T(Q^2) + a_n^{(4)} ~ \left[ {\alpha_s(Q^2) 
    \over \alpha_s(\mu^2)} \right]^{\gamma_n^{(4)}} ~ {\mu^2  \over Q^2} + 
    a_n^{(6)} ~ \left[ {\alpha_s(Q^2) \over  \alpha_s(\mu^2)} 
    \right]^{\gamma_n^{(6)}} ~ {\mu^4  \over Q^4}
    \label{eq:MnT}
 \ee
where the leading twist term $\mu_n^T(Q^2)$ is given at $NLO$ by Eq.
(\ref{eq:muT_NLO}) and $\mu = 1 ~ GeV/c$. All the five unknown parameters
(i.e., $a_n^{(2)}$, $a_n^{(4)}$, $\gamma_n^{(4)}$, $a_n^{(6)}$ and
$\gamma_n^{(6)}$) have been determined from the data through the
least-$\chi^2$ procedure for each value of $n$. The results are shown in
Figs. 6 and 7, while the values of the parameters are reported in Tables 2
and 3 in case of the proton and deuteron, respectively. Our main results can
be summarised as follows:

\begin{itemize}

\item our twist-2 term, extracted from the proton data together with the
twist-4 and twist-6 contributions, differs only slightly from the
predictions obtained using the set of parton distributions of Ref.
\cite{GRV} evolved at $NLO$ (see dashed lines in Fig. 6). We have checked
that, by repeating our analysis with the twist-2 term fixed at the $GRV$
prediction, the values of all the higher-twist parameters appearing in Eq.
(\ref{eq:MnT}) change for $n > 2$ only within the errors reported in Tables
2 and 3;

\item the twist-4 and twist-6 contributions result to have opposite signs and, moreover, at $Q^2 \sim 1 ~ (GeV/c)^2$ they are approximately of the same order of magnitude. The negative sign of the twist-6 is clearly due to the fact that a fit with a twist-4 term alone overestimates the low $Q^2$ data ($Q^2 \simeq$ few $(GeV/c)^2$). We stress that the opposite signs found for the twist-4 and twist-6 terms make the total higher-twist contribution smaller than its individual terms; in particular, at large $Q^2$ the sum of the twist-4 and twist-6 contributions turns out to be a small fraction of the twist-2 term ($\lsim 10 \%$ for $Q^2 \gsim n ~ (GeV/c)^2$).

\item the values of the effective anomalous dimensions $\gamma_n^{(4)}$ and $\gamma_n^{(6)}$ for $n = 4, 6, 8$ result to be around $4.0$ and $2.5$, respectively, i.e. significantly larger than the values of the corresponding twist-2 anomalous dimensions ($\gamma_n^{NS} \simeq 0.8 \div 1.2$ for $n = 4, 6, 8$ \cite{Altarelli});

\item the uncertainties on the different twist contributions due to the parameter fitting procedure are always within $\pm 15 \%$ (see Fig. 7 in case of the deuteron);

\item the twist expansions (\ref{eq:M2T}) and (\ref{eq:MnT}) appear to work quite well for values of $Q^2$ down to $\simeq 1 ~ (GeV/c)^2$.

\end{itemize}

\indent An interesting feature of our analysis of the transverse moments is that the leading-twist contribution is extracted from the analysis and not fixed by calculations based on a particular set of parton distributions. The comparison of our extracted twist-2 term with the predictions based on the $GRV$ parton distributions \cite{GRV} is shown in Fig. 8 for $Q^2 \gsim 5 ~ (GeV/c)^2$. It can clearly be seen that our results and the $GRV$ predictions agree quite well in case of the proton, whereas they differ significantly in case of the deuteron for $n > 2$. The inclusion of a new empirical determination of the nuclear effects in the deuteron, obtained in Ref. \cite{Gomez} from the nuclear dependence of the $SLAC$ data, increases only a little bit the disagreement for $n > 2$ (see dashed lines in Fig. 8). Since moments with $n > 2$ are mostly sensitive to the high-$x$ behaviour of the structure function, the question arises whether our extracted twist-2 moments can be explained by an enhancement of the $d$-quark parton distribution, like the one advocated in Ref. \cite{Bodek}, which explicitly reads as $\tilde{d}(x) = d(x) + 0.1 x (1 + x) u(x)$. The {\em modified} $GRV$ predictions, including also the empirical nuclear corrections for the deuteron, are shown by the solid lines in Fig. 8 and agree quite well with our results both for the proton and the deuteron. Therefore, our extracted twist-2 moments are clearly compatible with the hypothesis of an enhancement of the $d$-quark distribution at large $x$.

\indent So far the power corrections appearing in Eqs.
(\ref{eq:M2T}-\ref{eq:MnT}) have been derived assuming for the leading twist
the $NLO$ in the strong coupling constant. Since our main aim is to try to
estimate the target-dependent power corrections generated by multiparton
correlations, it is necessary to estimate the possible effects of higher
orders of the perturbative series, which defines the twist-2 coefficient
functions $E_{n2}(\mu, Q^2)$ appearing in Eq. (\ref{eq:CN}). However, it is
well known that such series is asymptotic at best and affected by the
so-called $IR$-renormalon ambiguities \cite{Beneke}. More precisely, the
only way to interpret consistently the large order behaviour of the
pertubartion theory leads unavoidably to ambiguities which have the general
form of power-suppressed terms. This happens because certain classes of
high-order radiative corrections to the twist-2 (the so-called fermion
bubble chains) are sensitive to large distances (i.e., to the
non-perturbative domain). Nevertheless it should be stressed that the
$IR$-renormalon contribution to the twist-2 has to cancel against the
ultraviolet quadratic divergencies of twist-4 operators (see, e.g., Refs.
\cite{Beneke,Balitsky,Stein}). This means that each twist in the expansion
(\ref{eq:CN}) is not defined unambigously, while the entire sum (i.e., the
complete calculation) is free from ambiguities. Therefore, while the data
cannot be affected by $IR$-renormalon uncertainties, our fitting procedure,
based on the separation among various twists and on the theoretical
perturbative treatment of the leading twist only, can suffer from
$IR$-renormalon ambiguities. The latter are target-independent quantities,
being of pure perturbative nature, while genuine higher-twist effects are
related to multiparton correlations in the target (i.e., target-dependent
quantities).

\indent The $IR$-renormalon picture has been applied to the phenomenology of
deep inelastic lepton-hadron scattering as a guide to estimate the $x$
dependence of power corrections \cite{Webber,Stein}. Such an estimate has
been found to be a good guess in case of the proton and the deuteron
structure functions and this fact may be understandable in terms of the
notion of a universal $IR$-finite effective strong coupling constant
\cite{DMW96} or in terms of the hypothesis of the dominance of the
quadratically divergent parts of the matrix elements of twist-4 operators
\cite{BBM97}.

\indent Before closing this Section, we point out that the $IR$-renormalon
contribution behaves as a power-like term, but it is characterised by
twist-2 anomalous dimensions. Thus, our observation that for $n = 4, 6, 8$
the effective anomalous dimensions $\gamma_n^{(4)}$ and $\gamma_n^{(6)}$
extracted from our $NLO$ fit to the transverse data are significantly
different from the corresponding twist-2 anomalous dimensions, might be an
indication of the presence of multiparton correlation effects (at least) in 
the transverse channel. In order to try to disentangle the latter from large
order perturbative effects, we will consider explicitly in the next Section
the power-like terms associated to the $IR$ renormalons as the general
uncertainty in the perturbative prediction of the twist-2 \cite{Braun}.

\section{$IR$ Renormalons and the Analysis of the Longitudinal Channel}

\indent Within the naive non-abelianization ($NNA$) approximation the contribution of the renormalon chains (i.e. the sum of vacuum polarisation insertions on the gluon line at one-loop level) to the non-singlet parts of the nucleon structure functions $F_1^N(x, Q^2)$ and $F_2^N(x, Q^2)$ is given by \cite{Webber}
 \be
    \label{eq:F1_IR}
    F_1^{IR}(x, Q^2) & = & \int_x^1 dz ~ F_1^{LT}({x \over z}, Q^2) ~ \left[ 
    A_2^{IR} {D_2(z) \over Q^2} + A_4^{IR} {D_4(z) \over Q^4} \right] \\
    \label{eq:F2_IR}
    F_2^{IR}(x, Q^2) & = & \int_x^1 dz ~ F_2^{LT}({x \over z}, Q^2) ~ \left[ 
    A_2^{IR} {C_2(z) \over Q^2} + A_4^{IR} {C_4(z) \over Q^4} \right]
 \ee
where the constants $A_2^{IR}$ and $A_4^{IR}$ are related to the log-moments of an $IR$-finite effective strong coupling constant \cite{DMW96}, $F_1^{LT}$ and $F_2^{LT}$ are the leading-twist structure functions and the coefficient functions $C_{2(4)}$ and $D_{2(4)}$ are given explicitly in Ref. \cite{Webber}. Thus, the $IR$-renormalon contribution to the transverse and longitudinal moments is explicitly given by
 \be
    \label{eq:muT_IR}
    \mu_n^{T(IR)}(Q^2) & = & \mu_n^{NS}(Q^2) \left\{ {A_2^{IR} \over Q^2} 
    \tilde{C}_2(n) + {A_4^{IR} \over Q^4} \tilde{C}_4(n) \right\} \\
    \label{eq:muL_IR} 
    \mu_n^{L(IR)}(Q^2) & = & \mu_n^{NS}(Q^2) \left\{ {A_2^{IR} \over Q^2}
    \left[ {8 \alpha_s(Q^2) \over 6 \pi} {\tilde{D}_2(n) \over n + 1} - 
    {4 n \over n + 2} \right] + \right . \nonumber \\
    & & \left. {A_4^{IR} \over Q^4} \left[ {8 \alpha_s(Q^2) \over 6 \pi} 
    {\tilde{D}_4(n) \over n + 1} - 4 n {n + 1 \over n + 3} \right] \right\}
 \ee
where in the last equation we have used the non-singlet part of the $NLO$ relation (\ref{eq:muL_NLO}). The coefficients $\tilde{C}_{2(4)}(n)$ and $\tilde{D}_{2(4)}(n)$ read as follows \cite{Webber}
 \be
    \label{eq:C_IR}
    \tilde{C}_2(n) & = & - n - 8 + {4 \over n} + {2 \over n + 1} + {12 \over 
    n + 2} + 4 S_n \\
    \tilde{C}_4(n) & = & {1 \over 2} n^2 - {3 \over 2} n + 16 - {4 \over n}
    - {4 \over n + 1} - {36 \over n + 3} - 4 S_n \nonumber \\
    \label{eq:D_IR}
    \tilde{D}_2(n) & = & - n - 4 + {4 \over n} + {2 \over n + 1} + {4 \over 
    n + 2} + 4 S_n \\
    \tilde{D}_4(n) & = & {1 \over 2} n^2 + {5 \over 2} n + 8 - {4 \over n}
    - {4 \over n + 1} - {12 \over n + 3} - 4 S_n \nonumber \\
 \ee
where $S_n \equiv \sum_{j = 1}^{n - 1} (1 / j)$. As already mentioned, Eqs. (\ref{eq:muT_IR}-\ref{eq:muL_IR}) describe the contributions of $IR$ renormalons to the non-singlet structure functions, while the more involved case of the singlet parts of the $DIS$ structure functions has been recently investigated in Ref. \cite{singlet}. There it has been shown that the difference between the $IR$-renormalon contributions to the singlet and non-singlet moments is not relevant for $n \geq 4$ thanks to the quark-gluon decoupling at large $x$. Therefore, for $n \geq 4$ it suffices to consider Eqs. (\ref{eq:muT_IR}-\ref{eq:muL_IR}) after substituting $\mu_n^{NS}(Q^2)$ with $\mu_n^T(Q^2)$, given at $NLO$ by Eq. (\ref{eq:muT_NLO}).

\indent An interesting feature of the $IR$-renormalon terms
(\ref{eq:muT_IR}-\ref{eq:muL_IR}) is that they are mainly governed by the
values of only two (unknown) parameters, $A_2^{IR}$ and $A_4^{IR}$, which
appear simultaneously both in the longitudinal and transverse channels at
any value of $n$. The signs of $A_2^{IR}$ and $A_4^{IR}$ are not
theoretically known, because they depend upon the prescription used to
circumvent the renormalon singularities of the Borel integrals, while within
the $NNA$ approximation their absolute values may be provided by \cite{Stein}  \be
    \label{eq:NNA}
    |A_2^{IR}| = {6 C_F \Lambda_{\overline{MS}}^2 \over 33 - 2 N_f} e^{5/3} 
    ~~, ~~~~~~~~~~~~
    |A_4^{IR}| = {3 C_F \Lambda_{\overline{MS}}^4 \over 33 - 2 N_f} e^{10/3}
 \ee
with $C_F = 4/3$. Using for $\Lambda_{\overline{MS}}$ the same values adopted for the $NLO$ calculation of $\alpha_s(Q^2)$ (see previous Section), one gets that $|A_2^{IR}|$ varies from $0.10$ to $0.13 ~ GeV^2$, while $|A_4^{IR}|$ ranges from $0.015$ to $0.030 ~ GeV^4$ for $N_f = 3, 4$.

\indent First of all, we have checked whether the power corrections to the
$NLO$ twist-2 contribution in the transverse channel can be explained by
pure $IR$-renormalon terms in the whole range $1 \lsim Q^2 ~ (GeV/c)^2 \lsim
20$. It turns out that: ~ i) the quality of the resulting fit is not as good
as the one obtained via the expansion (\ref{eq:MnT}) and the obtained minima
of the $\chi^2 / N$ variable can be much larger than $1$; ~ ii) the
extracted values of $|A_2^{IR}|$ and $|A_4^{IR}|$ result to be much greater
than the $NNA$ expectations (\ref{eq:NNA}) and to depend strongly upon the
inclusion of the data at low $Q^2$ ($Q^2 \sim$ few $(GeV/c)^2$). Moreover,
the $IR$-renormalon contribution to the longitudinal moments (Eq.
(\ref{eq:muL_IR})), calculated using the values of $A_2^{IR}$ and $A_4^{IR}$
determined from the analysis of the transverse channel, leads to a large
overestimation of the longitudinal data, as already noted in Ref.
\cite{Stein}. These results may be viewed as an effect of the presence of
higher-twist terms generated by multiparton correlations in the transverse
data for $Q^2 \gsim 1 ~ (GeV/c)^2$. To make this statement more
quantitative, we start with the analysis of the longitudinal data adopting
for the power corrections a pure $IR$-renormalon picture; namely, for $n
\geq 4$ we have used the following expansion
 \be
    M_n^L(Q^2) & = & \mu_n^L(Q^2) + \mu_n^{L(IR)}(Q^2) ~ \to_{n \geq 4} ~ 
    \mu_n^L(Q^2) + \mu_n^T(Q^2) \cdot \nonumber \\[3mm]
    & & \left\{ {A_2^{IR} \over Q^2} \left[ {8 \alpha_s(Q^2) \over 6 \pi} 
    {\tilde{D}_2(n) \over n + 1} - {4 n \over n + 2} \right] + {A_4^{IR} 
    \over Q^4} \left[ {8 \alpha_s(Q^2) \over 6 \pi} {\tilde{D}_4(n) \over n 
    + 1} - 4 n {n + 1 \over n + 3} \right] \right\} ~~~~
    \label{eq:MnL}
 \ee
where $\mu_n^L(Q^2)$ is given by Eq. (\ref{eq:muL_NLO}) and calculated using the $GRV$ parton distributions \cite{GRV}, while $\mu_n^T(Q^2)$ (see Eq. (\ref{eq:muT_NLO})) is taken from the $NLO$ analysis of the transverse data made in the previous Section (see Tables 2 and 3 for the values of the twist-2 parameters $a_n^{(2)}$).

\indent For each value of $n \geq 4$ we have determined the values of $A_2^{IR}$ and $A_4^{IR}$ from the least-$\chi^2$ fit to the longitudinal data in the whole range $1 \lsim Q^2 ~ (GeV/c)^2 \lsim 20$; our results are reported in Tables 4-5 and Figs. 9-10 in case of proton and deuteron targets, respectively. It can clearly be seen that: ~ i) the extracted values of $A_2^{IR}$ are almost independent of $n$, which means that the $n$-dependence (i.e., the shape in $x$) of the $1 / Q^2$ power correction is nicely predicted by the $NNA$ approximation; ~ ii) the values obtained for $|A_2^{IR}|$ are only slightly larger than the $NNA$ expectation (\ref{eq:NNA}); ~ iii) the determination of $A_4^{IR}$ is almost compatible with zero and affected by large uncertainties, since the $1 / Q^4$ power corrections turn out to be quite small in the longitudinal channel; nevertheless; the extracted values are not completely inconsistent with the $NNA$ predictions (\ref{eq:NNA}); ~ iv) the power corrections appear to be approximately the same in the proton and deuteron longitudinal channels (as it can be expected from a pure $IR$-renormalon phenomenology).

\indent As for the second moment $M_2^L(Q^2)$ we have simplified our analysis by taking only the non-singlet $IR$-renormalon contribution (\ref{eq:muL_IR}), i.e. by totally neglecting its singlet part. This is an approximation, but the resulting fit to the data on $M_2^L(Q^2)$ turns out to be quite good as it can be seen from Figs. 9(a) and 10(a), yielding values of $A_2^{IR}$ and $A_4^{IR}$ only slightly different from the ones previously determined by the analysis of the moments with $n \geq 4$ (see Tables 4 and 5).

\indent To sum up, in case of both proton and deuteron targets a pure
$IR$-renormalon description of power corrections works quite nicely in the
longitudinal channel starting already at $Q^2 \simeq 1 ~ (GeV/c)^2$.
Averaging the results of Tables 4 and 5 for $n \geq 4$ only, our
determination of the $IR$-renormalon strength parameters results to be:
$A_2^{IR} \simeq -0.132 \pm 0.015 ~ GeV^2$ and $A_4^{IR} \simeq 0.009 \pm
0.003 ~ GeV^4$, which, we stress, are not inconsistent with the $NNA$
expectations (\ref{eq:NNA}). Moreover, the value found for $A_2^{IR}$ nicely
agrees with the corresponding findings of Ref. \cite{Sidorov}, recently
obtained from a $NLO$ analysis of the $CCFR$ data \cite{CCFR} on $x F_3^N(x,
Q^2)$.

\indent Before closing this section, we stress again that the
$IR$-renormalon power-like terms should be regarded as the general
uncertainty in the perturbative calculation of the twist-2 term. Thus, it is
worth recalling that our estimates of the $IR$-renormalon parameters have
been obtained by taking the perturbative calculation of the twist-2 term at
$NLO$. We expect that our determination of the $IR$-renormalon parameters
holds only at $NLO$ and would vary if higher orders of the perturbation
theory were included. As a matter of fact, a significative reduction of the
$IR$-renormalon terms seems to be suggested by the recent $NNLO$ results
quoted in Refs. \cite{Bodek} and \cite{Sidorov}.   

\section{Final Analysis of the Transverse Data}

\indent After having determined the strengths and signs of the twist-4 and twist-6 $IR$-renormalon contributions from the analysis of the longitudinal channel, we can now proceed to the final analysis of the transverse data, adopting the following twist expansion
 \be
    \label{eq:MnT_final}
    M_n^T(Q^2) = \mu_n^T(Q^2) + \mu_n^{T(IR)}(Q^2) + a_n^{(4)} ~ \left[ 
    {\alpha_s(Q^2) \over \alpha_s(\mu^2)} \right]^{\gamma_n^{(4)}} ~ {\mu^2 
    \over Q^2} + a_n^{(6)} ~ \left[ {\alpha_s(Q^2) \over  \alpha_s(\mu^2)} 
    \right]^{\gamma_n^{(6)}} ~ {\mu^4  \over Q^4}
 \ee
where now the higher-twist terms involving the parameters $a_n^{(4)}$, $\gamma_n^{(4)}$, $a_n^{(6)}$ and $\gamma_n^{(6)}$ should be related to (target-dependent) multiparton correlation effects. Collecting Eqs. (\ref{eq:muT_NLO}) and (\ref{eq:muT_IR}) one has for $n \geq 4$
 \be
    M_{n \geq 4}^T(Q^2) & = & a_n^{(2)} \left[ {\alpha_s(Q^2) \over 
    \alpha_s(\mu^2)} \right]^{\gamma_n^{NS}} ~ {1 + \alpha_s(Q^2) R_n^{NS} 
    / 4\pi \over 1 + \alpha_s(\mu^2) R_n^{NS} / 4\pi} \left\{ 1 + {A_2^{IR}
    \over Q^2} \tilde{C}_2(n) + {A_4^{IR} \over Q^4} \tilde{C}_4(n) \right\} 
    + \nonumber \\
    & & a_n^{(4)} ~ \left[ {\alpha_s(Q^2) \over \alpha_s(\mu^2)} 
    \right]^{\gamma_n^{(4)}} ~ {\mu^2 \over Q^2} + a_n^{(6)} ~ \left[ 
    {\alpha_s(Q^2) \over  \alpha_s(\mu^2)} \right]^{\gamma_n^{(6)}} ~ {\mu^4 
    \over Q^4}
    \label{eq:MnT_n>2}
 \ee
The $IR$-renormalon parameters $A_2^{IR}$ and $A_4^{IR}$ have been kept fixed at the values $A_2^{IR} = -0.132 ~ GeV^2$ and $A_4^{IR} = 0.009 ~ GeV^4$ found in the previous Section, while the values of the five parameters $a_n^{(2)}$, $a_n^{(4)}$, $\gamma_n^{(4)}$, $a_n^{(6)}$ and $\gamma_n^{(6)}$ have been determined through the least-$\chi^2$ procedure and reported in Tables 6 and 7 in case of proton and deuteron, respectively. Comparing with the results obtained without the $IR$-renormalon contributions (see Tables 2 and 3), it can be clearly seen that the values of the twist-2 parameters $a_n^{(2)}$ are almost unchanged, while the values of the higher-twist parameters $a_n^{(4)}$, $\gamma_n^{(4)}$, $a_n^{(6)}$ and $\gamma_n^{(6)}$ vary only within the uncertainties of the fitting procedure. Note that with the inclusion of the $IR$-renormalon contribution the sum $a_n^{(4)} + a_n^{(6)}$ is closer to zero, which implies that at $Q^2 \simeq \mu^2 = 1 ~ (GeV/c)^2$ the twist-4 and twist-6 terms generated by multiparton correlation almost totally compensate each other. Such an effect is clearly illustrated in Figs. 11 and 12, where the contributions of the twist-2 at $NLO$, of the $IR$-renormalon and of the multiparton correlations are separately reported. From Figs. 11 and 12 it can be also seen that, for $n \geq 4$, the $IR$-renormalon contribution increases significantly around $Q^2 \sim 1 ~ (GeV/c)^2$ and could become of the same order of magnitude of the twist-2 term at $NLO$ in case of higher order moments. The effects from multiparton correlations appear to exceed the $IR$-renormalon term only for $Q^2 \gsim 2 ~ (GeV/c)^2$ (at $n \geq 4$). 

\indent As for the second moment $M_2^T(Q^2)$, following our previous analyses, we apply the $IR$-renormalon correction only to the non-singlet twist-2 term; moreover, the twist-2 parameters $a_2^S$ and $a_2^G$ are kept fixed at the values given in Table 1 for the deuteron and only the twist-4 term $a_n^{(4)} ~ \left[ \alpha_s(Q^2) / \alpha_s(\mu^2) \right]^{\gamma_n^{(4)}} ~ (\mu^2 / Q^2)$ is explicitly considered in the analysis. The resulting value of the twist-2 parameter $a_2^{NS}$ is $0.096 \pm 0.006$ ($0.066 \pm 0.0044$) for the proton (deuteron), which coincides within the uncertainties with the one given in Table 1. The twist-4 parameter $a_2^{(4)}$ turns out to be almost compatible with zero, namely $a_2^{(4)} = 0.01 \pm 0.01$ for the proton and $|a_2^{(4)}| \lsim 10^{-3}$ for the deuteron. Considering also the results of the previous Section on the longitudinal channel, our analyses indicate that the smallness of multiparton correlation effects on the second moments (both transverse and longitudinal ones) is consistent with the $Q^2$-behaviour of the data starting already at $Q^2 \simeq 1 ~ (GeV/c)^2$.  

\indent Basing on naive counting arguments (see, e.g., Ref. \cite{RGP77}), one can argue that the twist expansion for the transverse moments at $Q^2 \simeq \mu^2$ can be approximately rewritten as
 \be
    M_n^T(\mu^2) \simeq A_n^{(2)} \left[ 1 + n \left( {\Gamma_n^{(4)} \over 
    \mu} \right)^2 - n^2 \left( {\Gamma_n^{(6)} \over \mu} \right)^4 \right]
    \label{eq:mass_scale}
 \ee
where $A_n^{(2)}$ is the twist-2 contribution and $\Gamma_n^{(4)}$ ($\Gamma_n^{(6)}$) represents the mass scale of the twist-4 (twist-6) term, expected to be approximately independent of $n$ for $n \gsim 4$. (Note that in Eq. (\ref{eq:mass_scale}) we have already taken into account the opposite signs of the twist-4 and twist-6 terms as resulting from our analyses). Thus, one gets
 \be
    \label{eq:mass_ht}
    \Gamma_n^{(4)} = \mu \sqrt{{a_n^{(4)} \over n A_n^{(2)}}} ~~ , 
    ~~~~~~~~~~~~~~~~
    \Gamma_n^{(6)} = \mu \left[{|a_n^{(6)}| \over n^2 A_n^{(2)}} 
    \right]^{1/4}.
 \ee
Our results for $\Gamma_n^{(4)}$ and $\Gamma_n^{(6)}$ at $n = 4, 6, 8$,
obtained taking $a_n^{(4)}$ and $a_n^{(6)}$ from Tables 6 and 7 and using
for $A_n^{(2)}$ the twist-2 term at $NLO$ (i.e., $a_n^{(2)}$) {\em plus} the
whole $IR$-renormalon contribution (as determined from our fitting procedure
and evaluated at $Q^2 = \mu^2 = 1 ~ (GeV/c)^2$), are collected in Fig. 13.
It can clearly be seen that the mass scales of the twist-4 and twist-6 terms
are indeed approximately independent of $n$, viz. $\Gamma_n^{(4)} \simeq
\Gamma^{(4)} \simeq 0.76 ~ GeV$ and $\Gamma_n^{(6)} \simeq \Gamma^{(6)}
\simeq 0.55 ~ GeV$. The value obtained for $\Gamma^{(4)}$ is significantly
higher than the naive expectation $\Gamma^{(4)} \simeq \sqrt{<k_{\perp}^2>}
\simeq 0.3 ~ GeV$ \cite{RGP77,EFP83}, but not very far from the results of
Ref. \cite{Ji}. Without including the $IR$-renormalon contribution in
$A_n^{(2)}$ (i.e., taking only $A_n^{(2)} = a_n^{(2)}$), the values of
$\Gamma^{(4)}$ and $\Gamma^{(6)}$ would increase by $\simeq 20 \%$ and
$\simeq 10 \%$, respectively (cf. also Ref. \cite{JLAB98}).

\indent Before closing the Section, we want to address briefly the impact that the specific value adopted for the strong coupling constant at the $Z$-boson mass can have on our determination of multiparton correlation effects in the transverse channel. As already mentioned, existing analyses of the whole set of $DIS$ (unpolarised) world data favour the value $\alpha_s(M_Z^2) \simeq 0.113$ (see, e.g., Ref. \cite{Virchaux}), which is however well below the updated $PDG$ value $\alpha_s(M_Z^2) = 0.119 \pm 0.002$ \cite{PDG98}. Moreover, in Ref. \cite{Bodek} it has been argued that an increase of $\alpha_s(M_Z^2)$ up to $\simeq 0.120$ can give rise to a significative decrease of the relevance of the higher-twists effects in the $DIS$ data (up to a reduction by a factor $\simeq 2$). Therefore, we have repeated our analyses of longitudinal and transverse {\em pseudo-data} adopting the value $\alpha_s(M_Z^2) = 0.118$, where a set of parton distributions is available from Ref. \cite{MRS98} (we use the parton distributions only for the calculation of $\mu_n^L(Q^2)$ at $NLO$). All the results obtained at the higher value of $\alpha_s(M_Z^2)$ have the same quality as those presented at $\alpha_s(M_Z^2) = 0.113$ with a slight, but systematic increase of the minima of the $\chi^2 / N$ variable. The $IR$-renormalons parameters $A_2^{IR}$ and $A_4^{IR}$ are determined again by fitting the longitudinal data in the $Q^2$-range from $1$ to $20 ~ (GeV/c)^2$, and their values are now given by $A_2^{IR} \simeq -0.103 \pm 0.017$ and $A_4^{IR} \simeq 0.005 \pm 0.004$, which are compatible within the quoted uncertainties with the corresponding results of the high-$Q^2$ analysis of Ref. \cite{Bodek}. Thus, by construction, the sum of the twist-2 term at $NLO$ and the $IR$-renormalon contribution is almost independent of the specific value of $\alpha_s(M_Z^2)$ in the longitudinal channel. However, the same happens in the transverse channel as it is clearly illustrated in Fig. 14, where the results obtained at $\alpha_s(M_Z^2) = 0.113$ and $\alpha_s(M_Z^2) = 0.118$ are compared in case of the transverse moments with $n \geq 4$ and found to differ only by less than $\simeq 5 \%$. Therefore, though the $NLO$ twist-2 terms as well as the $IR$-renormalon contributions are separately sensitive to the specific value of $\alpha_s(M_Z^2)$, their sum turns out to be quite independent of $\alpha_s(M_Z^2)$. This means that our determination of the multiparton correlation effects is only marginally affected by the specific value adopted for $\alpha_s(M_Z^2)$, at least in the range of values from $\simeq 0.113$ to $\simeq 0.118$.

\section{Conclusions}

\indent We have analysed the power corrections to the $Q^2$ behaviour of the
low-order moments of both the longitudinal and transverse structure
functions of proton and deuteron using available phenomenological fits of
existing data in the $Q^2$ range between $1$ and $20 ~ (GeV/c)^2$. The
$SLAC$ proton data of Ref. \cite{Bosted}, which cover the region beyond $x
\simeq 0.7$ up to $x \simeq 0.98$, as well as existing data in the nucleon-resonance regions have been included in the analysis with the aim of
determining the effects of target-dependent higher-twists (i.e., multiparton
correlations).

\indent The Natchmann definition of the moments has been adopted for
disentangling properly kinematical target-mass and dynamical higher-twist
effects in the data. The leading twist has been treated at the $NLO$ in the
strong coupling constant and, as far as the transverse channel is concerned,
the twist-2 has been extracted simultaneously with the higher-twist terms.
The effects of higher orders of the perturbative series have been estimated
adopting the infrared renormalon model of Ref. \cite{Webber}, containing
both $1 / Q^2$ and $1 / Q^4$ power-like terms. It has been shown that the
longitudinal and transverse data cannot be explained simultaneously by the
renormalon contribution only; however, in the whole $Q^2$ range between $1$
and $20 ~ (GeV/c)^2$ the longitudinal channel appears to be consistent with
a pure $IR$-renormalon picture, adopting strengths not inconsistent with
those expected in the naive non-abelianization approximation and in nice
agreement with the results of a recent analysis \cite{Sidorov} of the $CCFR$
data \cite{CCFR} on $x F_3^N(x, Q^2)$. Then, after including the $1 / Q^2$
and $1 / Q^4$ $IR$-renormalon terms as fixed by the longitudinal channel,
the contributions of multiparton correlations to both the twist-4 and
twist-6 terms have been phenomenologically determined in the transverse
channel and found to have non-negligible strengths with opposite signs. It
has been also checked that our determination of the multiparton correlation
effects is only marginally affected by the specific value adopted for
$\alpha_s(M_Z^2)$ (at least in the range from $\simeq 0.113$ to $\simeq
0.118$).

\indent An interesting outcome of our analysis is that the extracted twist-2
contribution in the deuteron turns out to be compatible with the enhancement
of the $d$-quark parton distribution recently proposed in Ref. \cite{Bodek}.

\indent Let us stress that our analysis is presently limited by: ~ i) the
use of phenomenological fits of existing data (i.e., {\em pseudo-data}),
which are required in order to interpolate the structure functions in the
whole $x$-range for fixed values of $Q^2$, and ~ ii) by the existing large
uncertainties in the determination of the $L / T$ ratio $R_{L/T}^N(x, Q^2)$.
Therefore, both transverse data with better quality at $x \gsim 0.6$ and
$Q^2 \lsim 10 ~ (GeV/c)^2$ and more precise and systematic determinations of
the $L / T$ cross section ratio, which may be collected at planned
facilities like, e.g., $JLAB ~ @ ~ 12 ~ GeV$, could help to improve our
understanding of the non-perturbative structure of the nucleon. Finally, we
want to point out that, since in inclusive data multiparton correlations
appear to generate power-like terms with opposite signs, seminclusive or
exclusive experiments might offer the possibility to achieve a better
sensitivity to individual non-perturbative power corrections.

\section*{Acknowledgments}

\indent One of the author (S.S.) gratefully acknowledges Stefano Forte for many useful discussions about renormalons and power corrections during the progress of the paper.

\newpage

\centerline{\bf TABLES}

\vspace{1cm}

\noindent {\bf Table 1.} Values of the twist-2 parameters $a_2^{NS} \equiv
\mu_2^{NS}(\mu^2)$, $a_2^S \equiv \mu_2^S(\mu^2)$ and $a_2^G \equiv
\mu_2^G(\mu^2)$, and of the twist-4 parameters $a_2^{(4)}$ and
$\gamma_2^{(4)}$, obtained by fitting with Eq. (\ref{eq:M2T}) the
pseudo-data for the proton and the deuteron (see Figs. 2(a) and 3(a),
respectively). In case of the proton the parameters $a_2^S$ and $a_2^G$ have
been kept fixed at the values obtained for the deuteron. The last row
reports the value of the $\chi^2$ of the fit divided by the number of
degrees of freedom.  

\begin{table}[htb]

\begin{center}

{\bf Table 1}

\vspace{0.25cm}

\begin{tabular}{||c||c|c||}
\hline \hline
 $parm$ & $deuteron$ & $proton$
\\ \hline \hline
 $a_2^{NS}$       & $0.067 \pm 0.014$ & $0.092 \pm 0.007$
\\ \hline
 $a_2^S$          & $0.112 \pm 0.015$ & $0.112$
\\ \hline
 $a_2^G$          & $0.066 \pm 0.040$ & $0.066$
\\ \hline \hline
 $a_2^{(4)}$      & $\lsim 10^{-4}$   & $0.012 \pm 0.010$
\\ \hline
 $\gamma_2^{(4)}$ & ----              & $1 \pm 1$
\\ \hline \hline
 $\chi^2 / N$     & $0.11$            & $0.10$
\\ \hline \hline
\end{tabular}

\end{center}

\end{table}

\vspace{1cm}

\noindent {\bf Table 2.} Values of the twist-2 parameter $a_n^{(2)}$ and of the higher-twist parameters $a_n^{(4)}$, $\gamma_n^{(4)}$, $a_n^{(6)}$ and $\gamma_n^{(6)}$, obtained by fitting with Eq. (\ref{eq:MnT}) the proton pseudo-data of Fig. 2 for $n = 4, 6$ and $8$. The last row reports the value of the $\chi^2$ of the fit divided by the number of degrees of freedom.

\begin{table}[htb]

\begin{center}

{\bf Table 2}

\vspace{0.25cm}

\begin{tabular}{||c||c|c|c||}
\hline \hline
 $parm$ & \multicolumn{3}{c||}{$proton$} \\ \cline{2-4}
 & $M_4^T$ & $M_6^T$ & $M_8^T$
\\ \hline \hline
 $a_n^{(2)}$      & $0.028 \pm 0.001$  & $0.0078 \pm 0.0003$ & $0.0030 \pm 
0.0001$ \\ \hline \hline
 $a_n^{(4)}$      & $0.071 \pm 0.009$  & $0.046 \pm 0.003$   & $0.029 \pm 
0.003$ \\ \hline
 $\gamma_n^{(4)}$ & $3.8 \pm 0.4$      & $3.7 \pm 0.2$       & $3.6 \pm 0.2$ 
\\ \hline \hline
 $a_n^{(6)}$      & $-0.056 \pm 0.008$ & $-0.039 \pm 0.005$  & $-0.026 \pm 
0.002$ \\ \hline
 $\gamma_n^{(6)}$ & $2.5 \pm 0.5$      & $2.3 \pm 0.2$       & $2.1 \pm 0.4$ 
\\ \hline \hline
 $\chi^2 / N$     & $0.05$             & $0.27$              & $0.65$ \\
\hline \hline
\end{tabular}

\end{center}

\end{table}

\newpage

\vspace{1cm}

\noindent {\bf Table 3.} The same as in Table 2, but for the deuteron.

\begin{table}[htb]

\begin{center}

{\bf Table 3}

\vspace{0.25cm}

\begin{tabular}{||c||c|c|c||}
\hline \hline
 $parm$ & \multicolumn{3}{c||}{$deuteron$} \\ \cline{2-4}
 & $M_4^T$ & $M_6^T$ & $M_8^T$
\\ \hline \hline
 $a_n^{(2)}$      & $0.022 \pm 0.001$  & $0.0060 \pm 0.0002$ & $0.0024 \pm
0.0001$ \\ \hline \hline
 $a_n^{(4)}$      & $0.055 \pm 0.006$  & $0.036 \pm 0.004$   & $0.022 \pm 
0.002$ \\ \hline
 $\gamma_n^{(4)}$ & $3.9 \pm 0.2$      & $4.0 \pm 0.2$       & $3.8 \pm 0.3$ 
\\ \hline \hline
 $a_n^{(6)}$      & $-0.046 \pm 0.005$ & $-0.032 \pm 0.004$  & $-0.020 \pm
0.002$ \\ \hline
 $\gamma_n^{(6)}$ & $2.5 \pm 0.2$      & $2.4 \pm 0.2$       & $2.2 \pm 0.5$ 
\\ \hline \hline
 $\chi^2 / N$     & $0.10$             & $0.37$               & $1.0$ \\
\hline \hline
\end{tabular}

\end{center}

\end{table}

\vspace{1cm}

\noindent {\bf Table 4.} Values of the $IR$-renormalon parameters $A_2^{IR}$ and $A_4^{IR}$ obtained using Eq. (\ref{eq:MnL}) to fit the longitudinal proton data of Fig. 4. The twist-2 contribution $\mu_n^L(Q^2)$ is calculated at $NLO$ via Eq. (\ref{eq:muL_NLO}) adopting the $GRV$ set of parton distributions \cite{GRV}, while $\mu_n^T(Q^2)$ (see Eq. (\ref{eq:muT_NLO})) is obtained using the values of $a_n^{(2)}$ reported in Table 2. The last row reports the value of the $\chi^2$ of the fit divided by the number of degrees of freedom.

\begin{table}[htb]

\begin{center}

{\bf Table 4}

\vspace{0.25cm}

\begin{tabular}{||c||c|c|c|c||}
\hline \hline
 $parm$ & \multicolumn{4}{c||}{$proton$} \\ \cline{2-5}
 & $M_2^L$ & $M_4^L$ & $M_6^L$ & $M_8^L$
\\ \hline \hline
 $A_2^{IR}$   & $-0.16 \pm 0.05$ & $-0.12 \pm 0.04$ & $-0.11 \pm 0.04$   & 
$-0.13 \pm 0.04$ \\ \hline
 $A_4^{IR}$   & $0.06 \pm 0.03$  & $0.01 \pm 0.01$  & $-0.001 \pm 0.001$ & 
$0.0004 \pm 0.0004$ \\ \hline \hline
 $\chi^2 / N$ & $0.24$           & $1.5$            & $1.1$              & 
$0.42$ \\ \hline \hline
\end{tabular}

\end{center}

\end{table}

\newpage

\vspace{1cm}

\noindent {\bf Table 5.} The same as in Table 4, but for the deuteron.

\begin{table}[htb]

\begin{center}

{\bf Table 5}

\vspace{0.25cm}

\begin{tabular}{||c||c|c|c|c||}
\hline \hline
 $parm$ & \multicolumn{4}{c||}{$deuteron$} \\ \cline{2-5}
 & $M_2^L$ & $M_4^L$ & $M_6^L$ & $M_8^L$
\\ \hline \hline
 $A_2^{IR}$   & $-0.18 \pm 0.06$ & $-0.14 \pm 0.04$  & $-0.14 \pm 0.04$  & 
$-0.15 \pm 0.01$ \\ \hline
 $A_4^{IR}$   & $0.06 \pm 0.03$  & $0.025 \pm 0.013$ & $0.012 \pm 0.008$ & 
$0.009 \pm 0.003$ \\ \hline \hline
 $\chi^2 / N$ & $0.09$           & $0.26$            & $0.36$            & 
$0.18$ \\ \hline \hline
\end{tabular}

\end{center}

\end{table}

\vspace{1cm}

\noindent {\bf Table 6.} Values of the twist-2 parameter $a_n^{(2)}$ and of the higher-twist parameters $a_n^{(4)}$, $\gamma_n^{(4)}$, $a_n^{(6)}$ and $\gamma_n^{(6)}$, obtained by fitting with Eq. (\ref{eq:MnT_n>2}) the proton pseudo-data of Fig. 2 for $n = 4, 6$ and $8$. The $IR$-renormalon twist-4 and twist-6 contributions are obtained using $A_2^{IR} = -0.132 ~ GeV^2$ and $A_4^{IR} = 0.009 ~ GeV^4$ (see text). The last row reports the value of the $\chi^2$ of the fit divided by the number of degrees of freedom.

\begin{table}[htb]

\begin{center}

{\bf Table 6}

\vspace{0.25cm}

\begin{tabular}{||c||c|c|c||}
\hline \hline
 $parm$ & \multicolumn{3}{c||}{$proton$} \\ \cline{2-4}
 & $M_4^T$ & $M_6^T$ & $M_8^T$
\\ \hline \hline
 $a_n^{(2)}$      & $0.028 \pm 0.001$  & $0.0079 \pm 0.0003$ & $0.0031 \pm 
0.0001$ \\ \hline \hline
 $a_n^{(4)}$      & $0.061 \pm 0.008$  & $0.045 \pm 0.002$   & $0.028 \pm 
0.003$ \\ \hline
 $\gamma_n^{(4)}$ & $4.1 \pm 0.2$      & $4.1 \pm 0.2$       & $4.1 \pm 0.5$
\\ \hline \hline
 $a_n^{(6)}$      & $-0.051 \pm 0.008$ & $-0.041 \pm 0.002$  & $-0.027 \pm 
0.003$ \\ \hline
 $\gamma_n^{(6)}$ & $2.7 \pm 0.2$      & $2.6 \pm 0.2$       & $2.4 \pm 0.2$ \\ \hline \hline
 $\chi^2 / N$     & $0.06$             & $0.21$              & $0.66$ \\
\hline \hline
\end{tabular}

\end{center}

\end{table}

\newpage 

\vspace{1cm}

\noindent {\bf Table 7.} The same as in Table 6, but for the deuteron.

\begin{table}[htb]

\begin{center}

{\bf Table 7}

\vspace{0.25cm}

\begin{tabular}{||c||c|c|c||}
\hline \hline
 $parm$ & \multicolumn{3}{c||}{$deuteron$} \\ \cline{2-4}
 & $M_4^T$ & $M_6^T$ & $M_8^T$
\\ \hline \hline
 $a_n^{(2)}$      & $0.022 \pm 0.001$  & $0.0060 \pm 0.0002$ & $0.0024 \pm 
0.0001$ \\ \hline \hline
 $a_n^{(4)}$      & $0.047 \pm 0.004$  & $0.032 \pm 0.003$   & $0.020 \pm 
0.002$ \\ \hline
 $\gamma_n^{(4)}$ & $4.4 \pm 0.2$      & $4.2 \pm 0.2$       & $4.1 \pm 0.3$ 
\\ \hline \hline
 $a_n^{(6)}$      & $-0.043 \pm 0.003$ & $-0.030 \pm 0.003$  & $-0.019 \pm 
0.002$ \\ \hline
 $\gamma_n^{(6)}$ & $2.9 \pm 0.2$      & $2.6 \pm 0.2$       & $2.4 \pm 0.3$ 
\\ \hline \hline
 $\chi^2 / N$     & $0.10$             & $0.37$              & $1.0$ \\
\hline \hline
\end{tabular}

\end{center}

\end{table}

\newpage

\centerline{\Large \bf FIGURES}

\vspace{4cm}

\begin{figure}[htb]

\centerline{\Large{\bf DEUTERON}} \vspace{0.5cm}

\centerline{\epsfxsize=14cm \epsfig{file=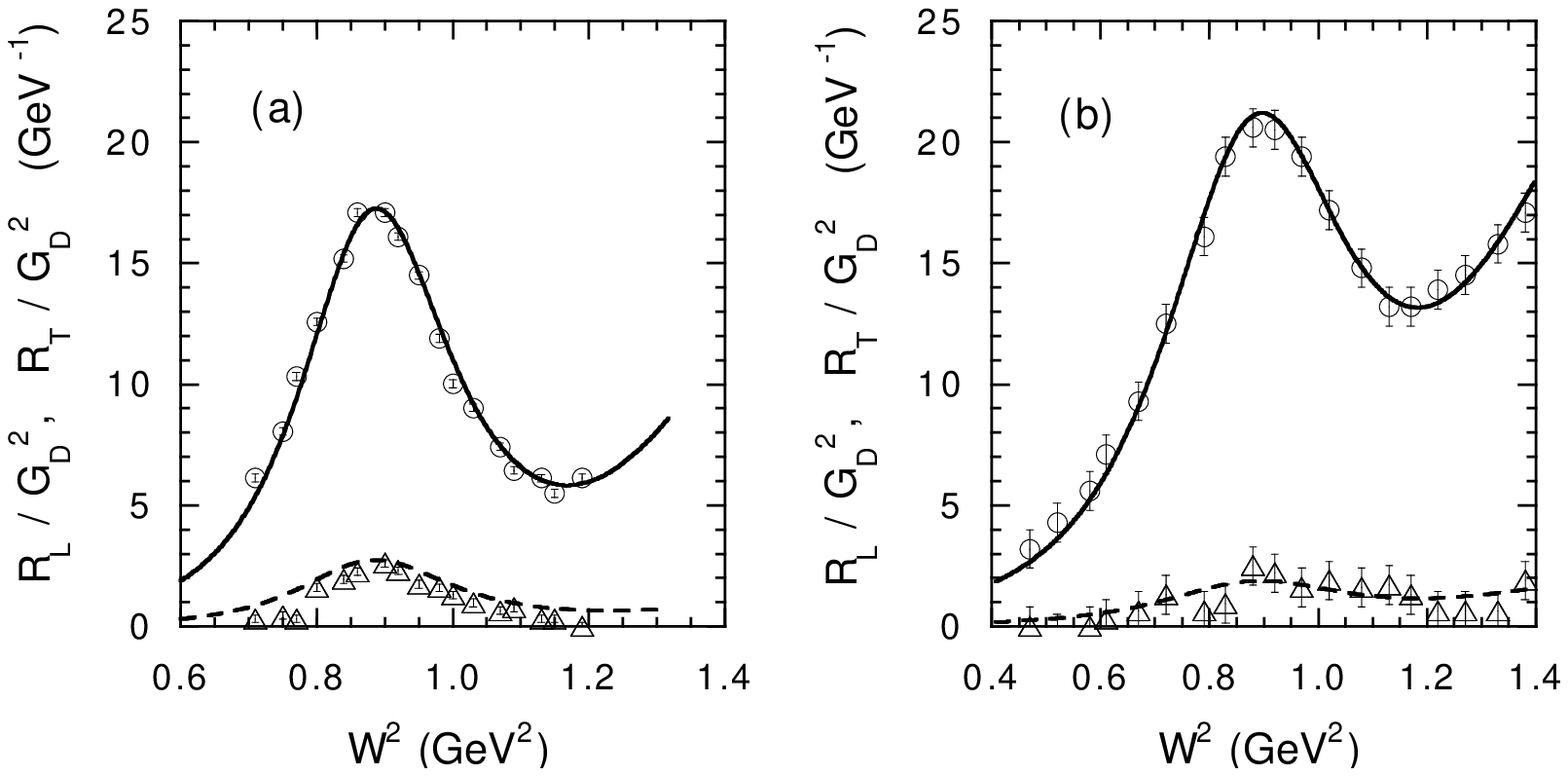}}

\rightline{} \vspace{1cm}

\small{{\bf Figure 1.} Reduced response functions of the deuteron versus the squared invariant mass $W^2 \equiv M^2 + Q^2 \cdot (1 / x - 1)$, measured at $SLAC$ \cite{SLAC_deut} for $Q^2 = 1.75 ~ (GeV/c)^2$ (a) and $Q^2 = 3.25 ~ (GeV/c)^2$ (b). The open dots and triangles correspond to the (reduced) longitudinal $R_L$ and transverse $R_T$ responses, respectively. The solid and dashed lines are the results obtained using the procedure of Ref. \cite{CS}. The dipole form factor $G_D(Q^2)$ is explicitly given by $G_D = 1 / (1 + Q^2 / 0.71)^2$.}

\end{figure}

\newpage

\begin{figure}[htb]

\centerline{\Large{\bf PROTON}} \vspace{0.5cm}

\centerline{\epsfxsize=14cm \epsfig{file=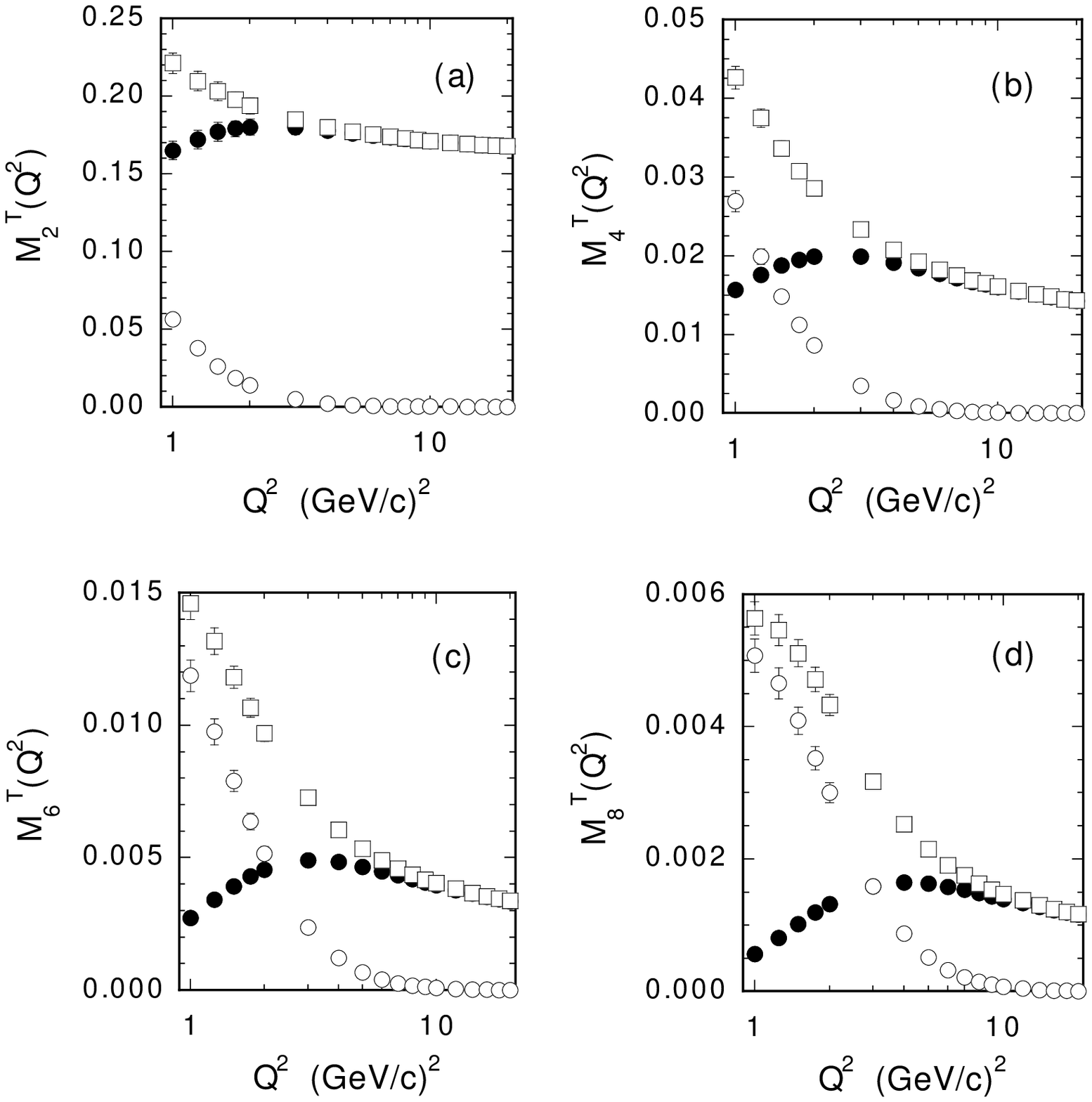}}

\rightline{} \vspace{1cm}

\small{{\bf Figure 2.} Pseudo-data for the (Natchmann) transverse moments $M_n^T(Q^2)$ (Eq. (\ref{eq:MT})) of the proton versus $Q^2$ for $n = 2, 4, 6$ and $8$. Open and full dots correspond to the inelastic and elastic contributions, respectively, while the open squares are the total moments. The elastic contribution is evaluated via Eq. (\ref{eq:MTel}) assuming a dipole form for the form factors.}

\end{figure}

\newpage

\begin{figure}[htb]

\centerline{\Large{\bf DEUTERON}} \vspace{0.5cm}

\centerline{\epsfxsize=14cm \epsfig{file=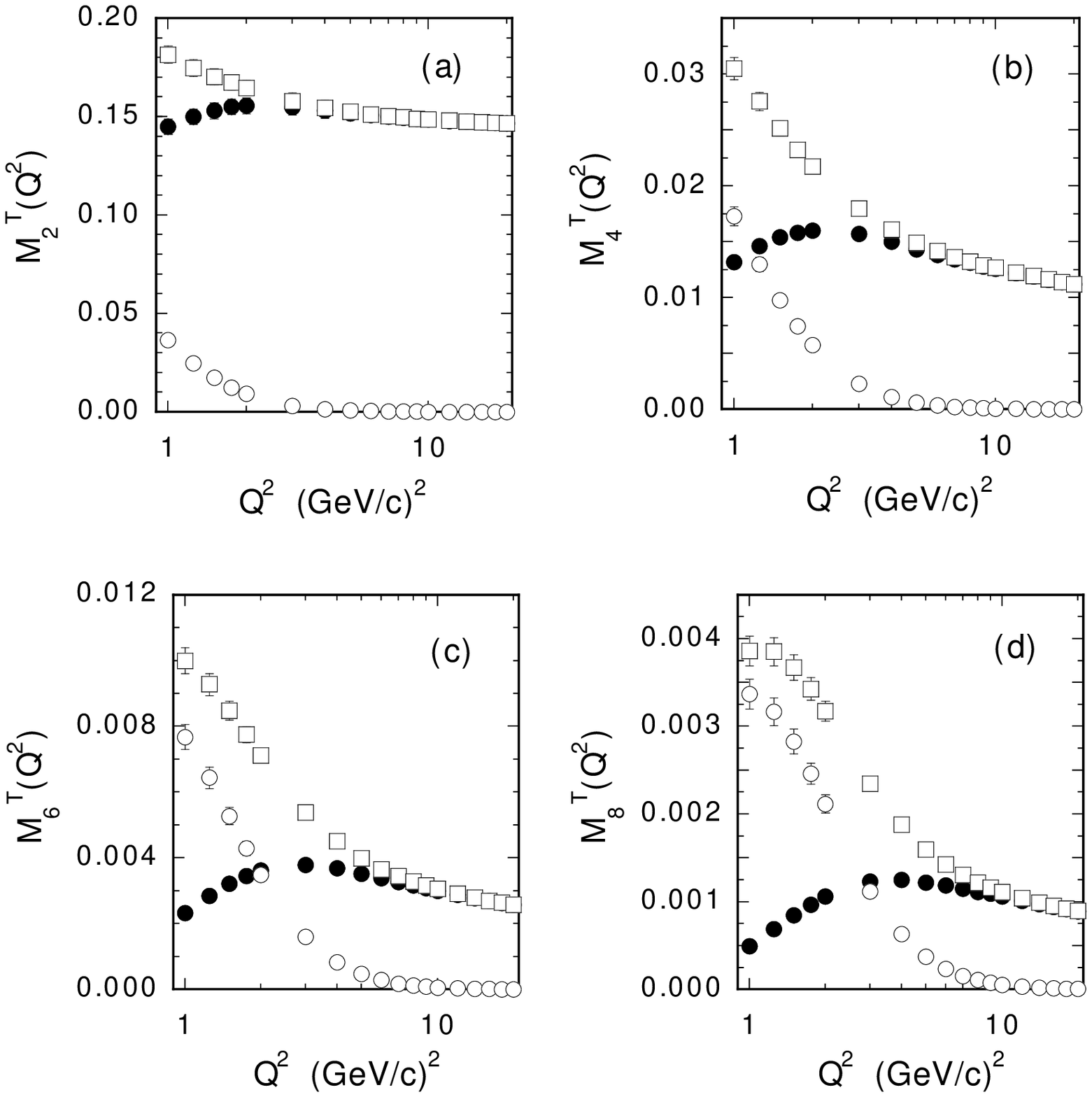}}

\rightline{} \vspace{1cm}

\centerline{\small{{\bf Figure 3.} The same as in Fig. 2, but for the deuteron.}}

\end{figure}

\newpage

\begin{figure}[htb]

\centerline{\Large{\bf PROTON}} \vspace{0.5cm}

\centerline{\epsfxsize=14cm \epsfig{file=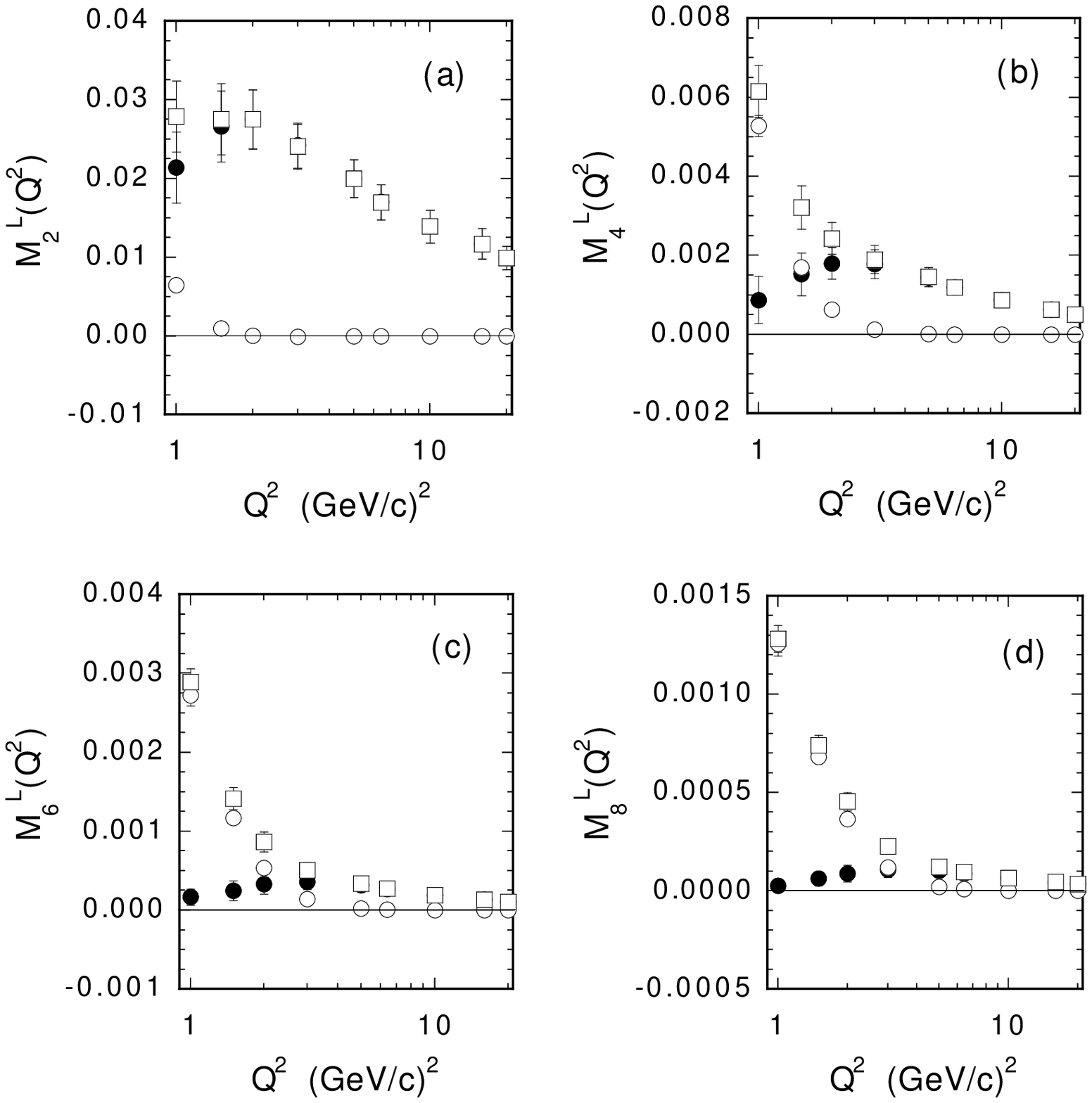}}

\rightline{} \vspace{1cm}

\small{{\bf Figure 4.} Pseudo-data for the (Natchmann) longitudinal moments $M_n^L(Q^2)$ (Eq. (\ref{eq:ML})) of the proton versus $Q^2$ for $n = 2, 4, 6$ and $8$. Open and full dots correspond to the inelastic and elastic contributions, respectively, while the open squares are the total moments. The elastic contribution is evaluated via Eq. (\ref{eq:MLel}) assuming a dipole form for the form factors.}

\end{figure}

\newpage

\begin{figure}[htb]

\centerline{\Large{\bf DEUTERON}} \vspace{0.5cm}

\centerline{\epsfxsize=14cm \epsfig{file=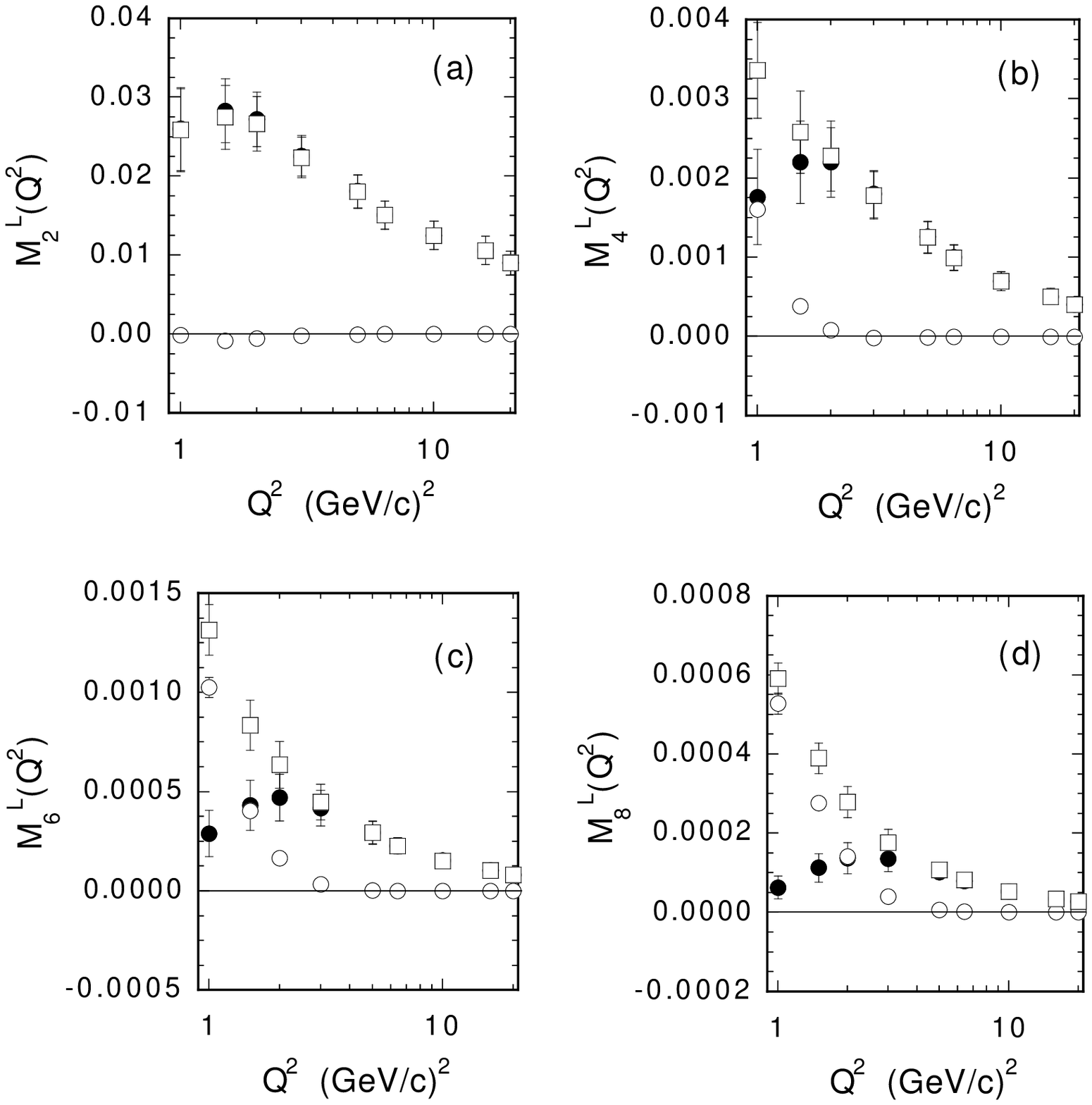}}

\rightline{} \vspace{1cm}

\centerline{\small{{\bf Figure 5.} The same as in Fig. 4, but for the deuteron.}}

\end{figure}

\newpage

\begin{figure}[htb]

\centerline{\Large{\bf PROTON}} \vspace{0.5cm}

\centerline{\epsfxsize=14cm \epsfig{file=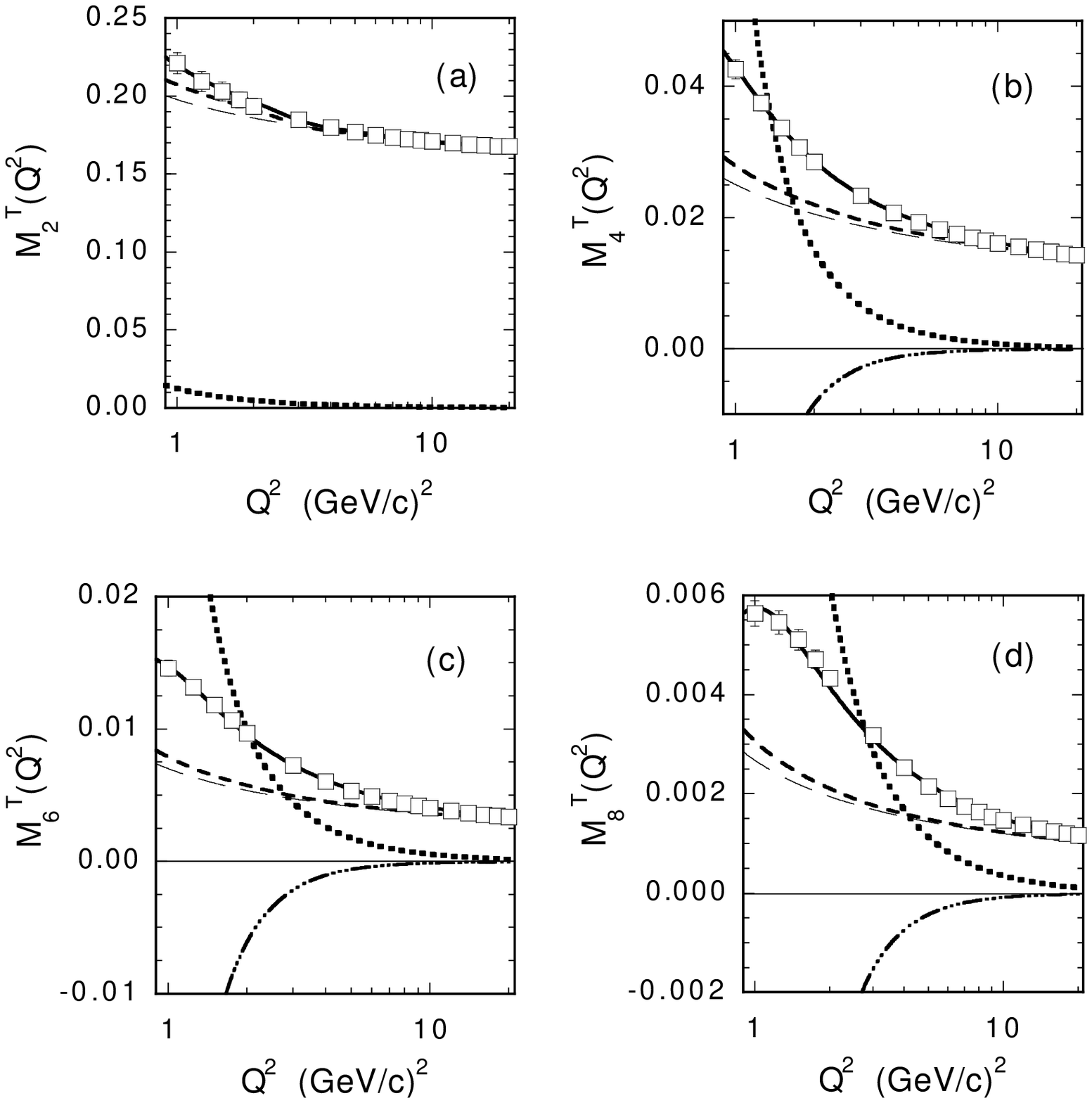}}

\rightline{} \vspace{1cm}

\small{{\bf Figure 6.} Twist analysis of the proton transverse moments $M_n^T(Q^2)$ (Eq. (\ref{eq:MT})) for $n = 2, 4, 6$ and $8$. The open squares represent the pseudo-data of Fig. 2. The thick solid lines are the result of Eqs. (\ref{eq:M2T}) and (\ref{eq:MnT}) fitted by the least-$\chi^2$ procedure to the pseudo-data of Fig. 2. The thick dashed, dotted and dot-dashed lines correspond to the contributions of the twist-2, twist-4 and twist-6, respectively. The thin long-dashed lines are the predictions of the twist-2 term obtained using the parton distributions of Ref. \cite{GRV}.}

\end{figure}

\newpage

\begin{figure}[htb]

\centerline{\Large{\bf DEUTERON}} \vspace{0.5cm}

\centerline{\epsfxsize=14cm \epsfig{file=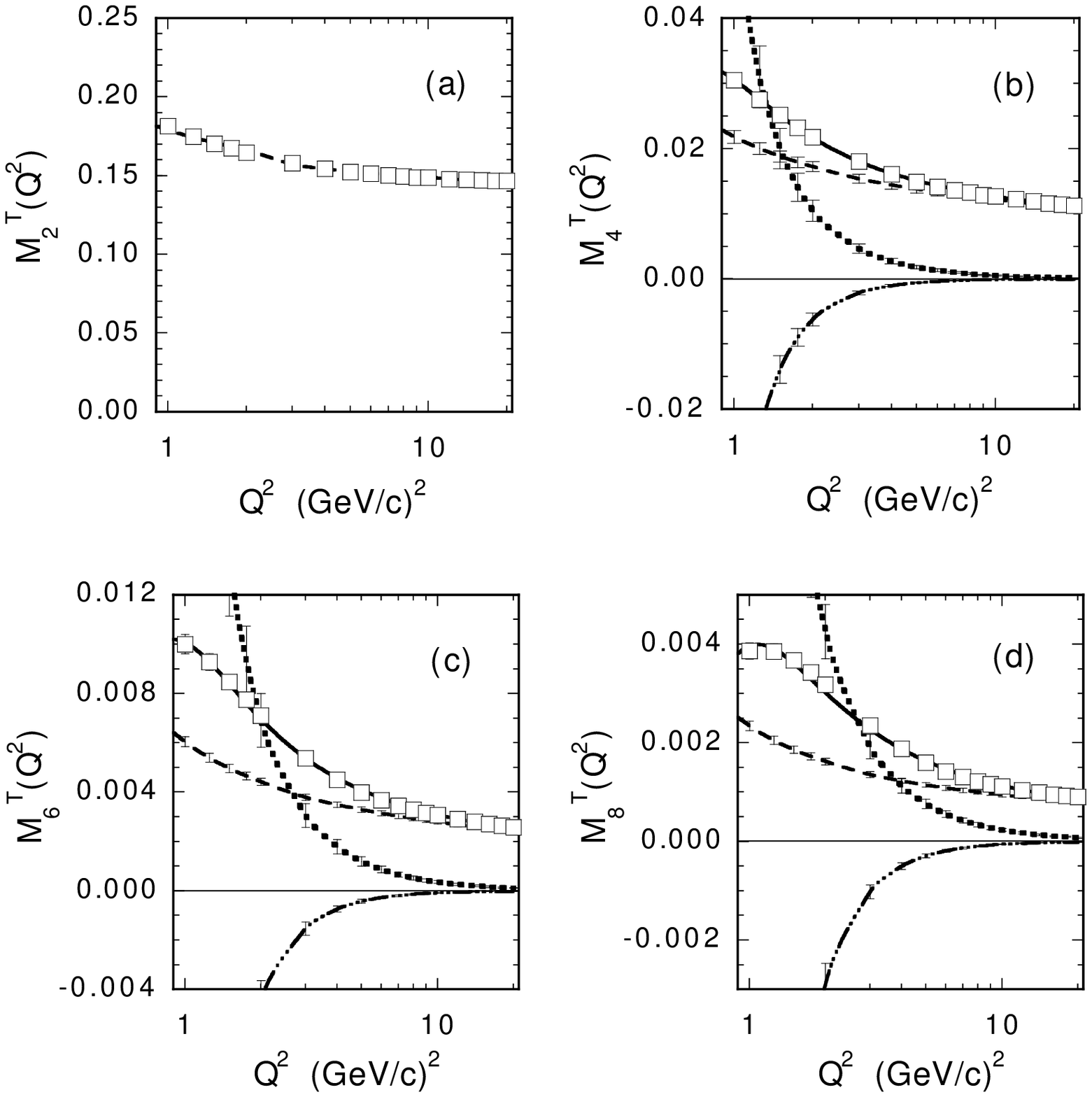}}

\rightline{} \vspace{1cm}

\small{{\bf Figure 7.} The same as in  Fig. 6, but for the deuteron. In (a) only the twist-2 term is reported (see text). In (b), (c) and (d) the error bars on the different twist contributions correspond to the total uncertainties generated by the errors of the parameters appearing in Table 3.}

\end{figure}

\newpage

\begin{figure}[htb]

\centerline{\Large ~~~~~~~~~~~~ {\bf PROTON} ~~~~~~~~~~~~~~~~~~~~~ {\bf DEUTERON}} \vspace{0.5cm}

\centerline{\epsfxsize=14cm \epsfig{file=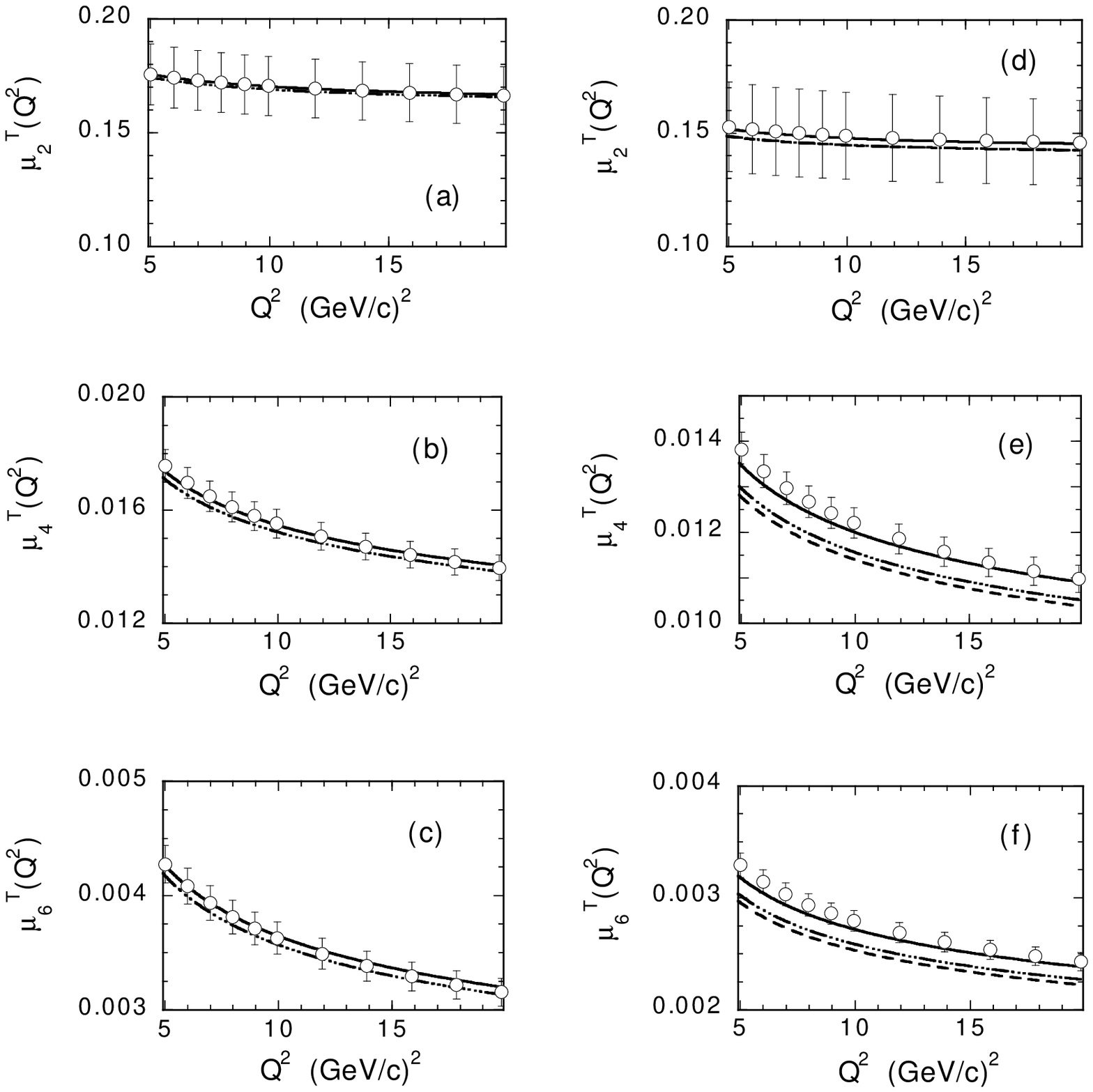}}

\rightline{} \vspace{1cm}

\small{{\bf Figure 8.} The leading-twist moments $\mu_n^T(Q^2)$ (see Eqs. (\ref{eq:muT}-\ref{eq:muS_NLO})) versus $Q^2$ for $n = 2, 4, 6$. Open dots: twist-2 extracted from our analysis of the proton and deuteron transverse pseudo-data (see also Tables 1-3); the errors represent the uncertainty of the fitting procedure corresponding to one-unit increment of the $\chi^2 / N$ variable. Dot-dashed lines: $GRV$ prediction at $NLO$ \cite{GRV}. Dashed lines: the same as the dot-dashed lines, but including the empirical nuclear correction of Ref. \cite{Gomez}. Solid lines: the same as the dashed lines, but including the enhancement of the $d$-quark distribution at large $x$ of Ref. \cite{Bodek}, given explicitly by $\tilde{d}(x) = d(x) + 0.1 x (1 + x) u(x)$.}

\end{figure}

\newpage

\begin{figure}[htb]

\centerline{\Large{\bf PROTON}} \vspace{0.5cm}

\centerline{\epsfxsize=14cm \epsfig{file=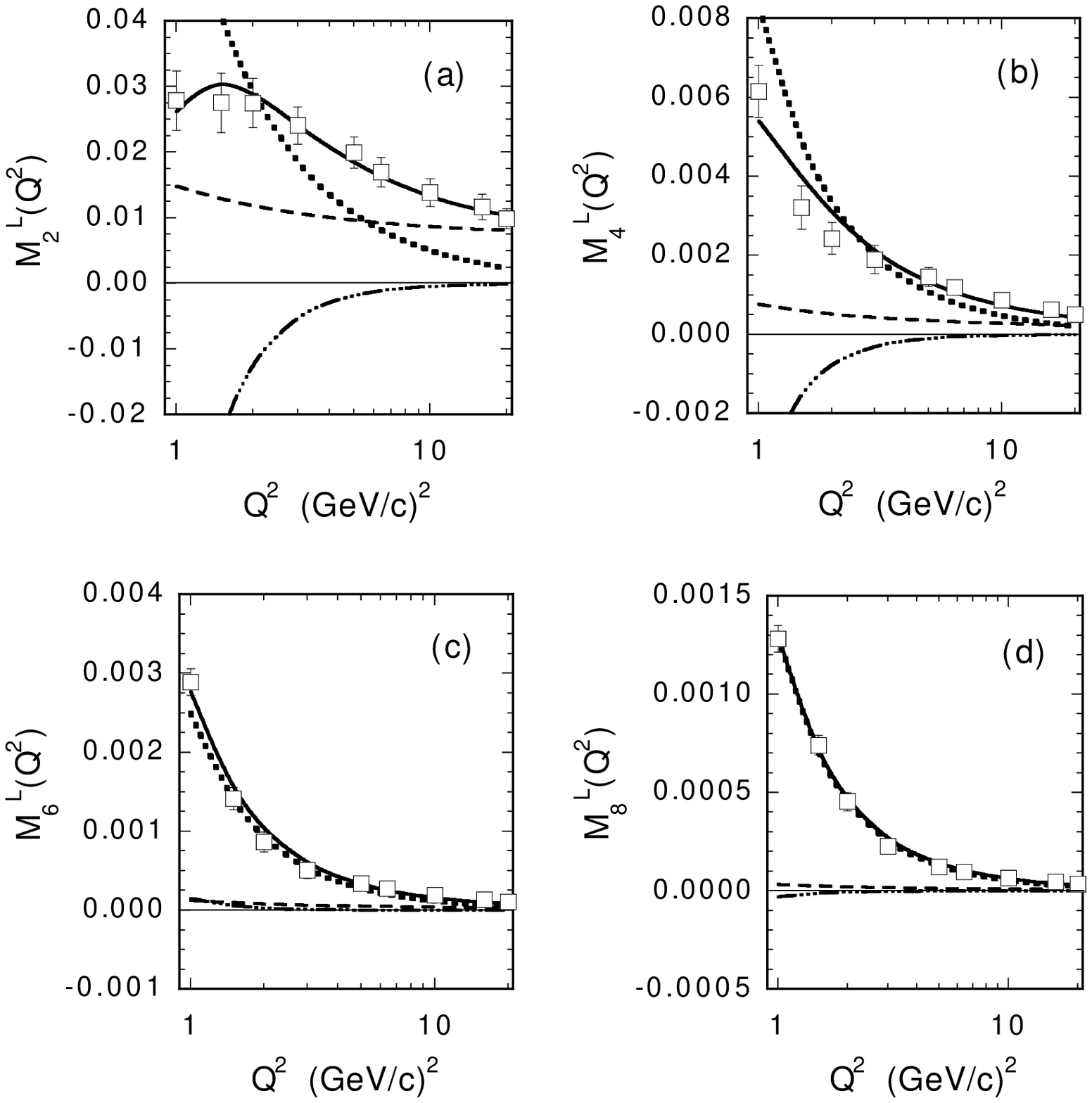}}

\rightline{} \vspace{1cm}

\small{{\bf Figure 9.} Twist analysis of the proton longitudinal moments
$M_n^L(Q^2)$ (Eq. (\ref{eq:ML})) for $n = 2, 4, 6$ and $8$, adopting the
$IR$-renormalon model of Ref. \cite{Webber}. The open squares represent the
pseudo-data of Fig. 7. The solid lines are the results of Eq. (\ref{eq:MnL}) 
fitted by the least-$\chi^2$ procedure to the pseudo-data. The dashed lines 
are the twist-2 contribution given by Eq. (\ref{eq:muL_NLO}) adopting the 
$GRV$ set of parton distributions \cite{GRV} at $NLO$. The dotted and 
dot-dashed lines correspond to the twist-4 and twist-6 terms, respectively.}

\end{figure}

\newpage

\begin{figure}[htb]

\centerline{\Large{\bf DEUTERON}} \vspace{0.5cm}

\centerline{\epsfxsize=14cm \epsfig{file=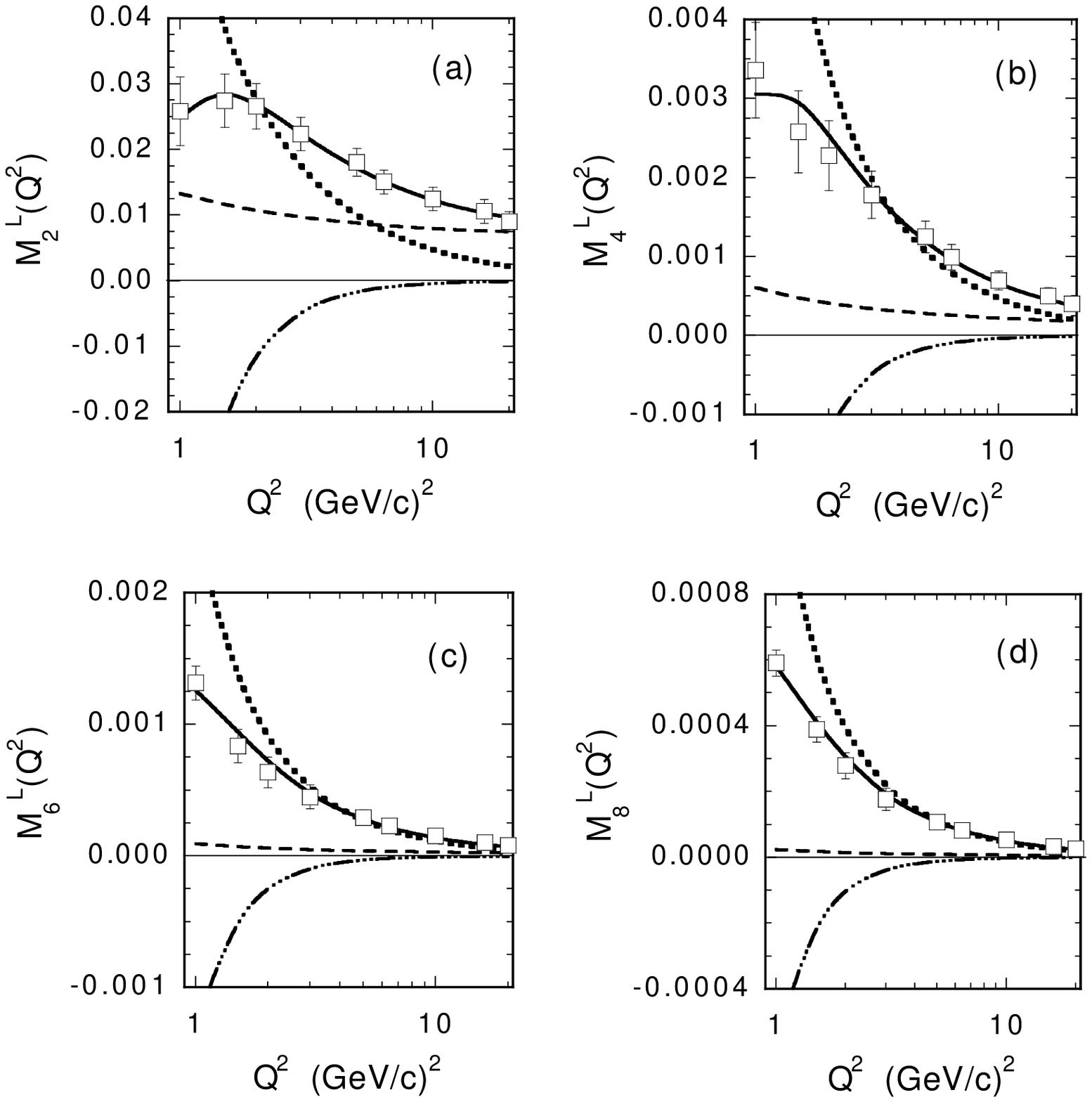}}

\rightline{} \vspace{1cm}

\centerline{\small{{\bf Figure 10.} The same as in  Fig. 9, but for the deuteron.}}

\end{figure}

\newpage

\begin{figure}[htb]

\centerline{\Large{\bf PROTON}} \vspace{0.5cm}

\centerline{\epsfxsize=14cm \epsfig{file=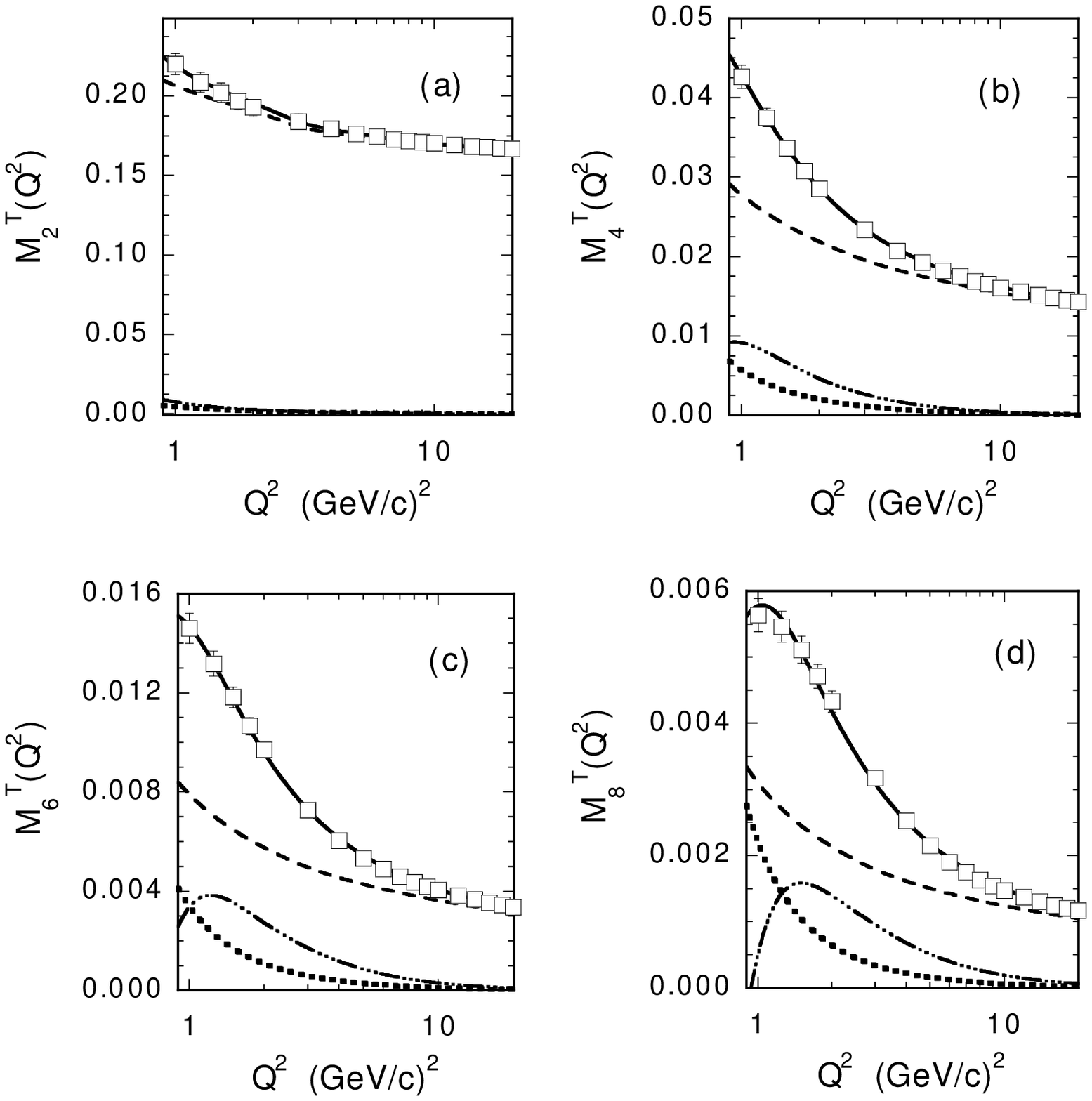}}

\rightline{} \vspace{1cm}

\small{{\bf Figure 11.} Twist analysis of the proton transverse moments $M_n^T(Q^2)$ (Eq. (\ref{eq:MT})) for $n = 2, 4, 6$ and $8$, including both phenomenological higher twists originating from multiparton correlations and the power corrections arising from the $IR$-renormalon model of Ref. \cite{Webber} (see text). The values of the $IR$-renormalon parameters are fixed at $A_2^{IR} = -0.132 ~ GeV^2$ and $A_4^{IR} = 0.009 ~ GeV^4$, as resulting from the analysis of the longitudinal moments. The solid lines are the results of Eq. (\ref{eq:MnT_final}) fitted by the least-$\chi^2$ procedure to the pseudo-data (open squares). The dashed lines are the twist-2 contribution, while the dotted and dot-dashed lines correspond to the $IR$-renormalon and multiparton correlation contributions, respectively.}

\end{figure}

\newpage

\begin{figure}[htb]

\centerline{\Large{\bf DEUTERON}} \vspace{0.5cm}

\centerline{\epsfxsize=14cm \epsfig{file=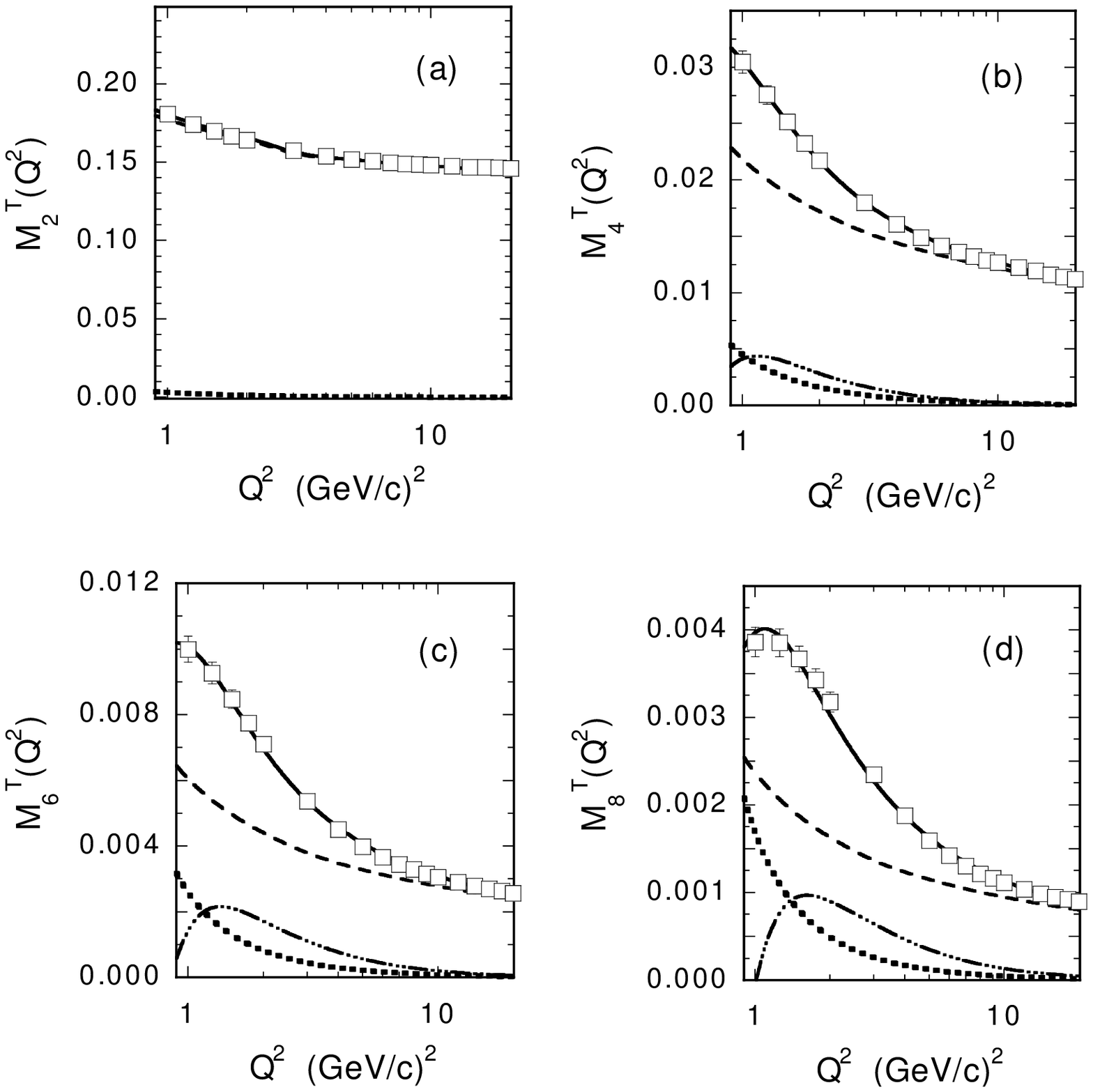}}

\rightline{} \vspace{1cm}

\small{{\bf Figure 12.} The same as in  Fig. 11, but for the deuteron. In (a) only the twist-2 term and the $IR$-renormalon contribution are reported (see text).}

\end{figure}

\newpage

\begin{figure}[htb]

\rightline{} \vspace{2cm}

\centerline{\epsfxsize=14cm \epsfig{file=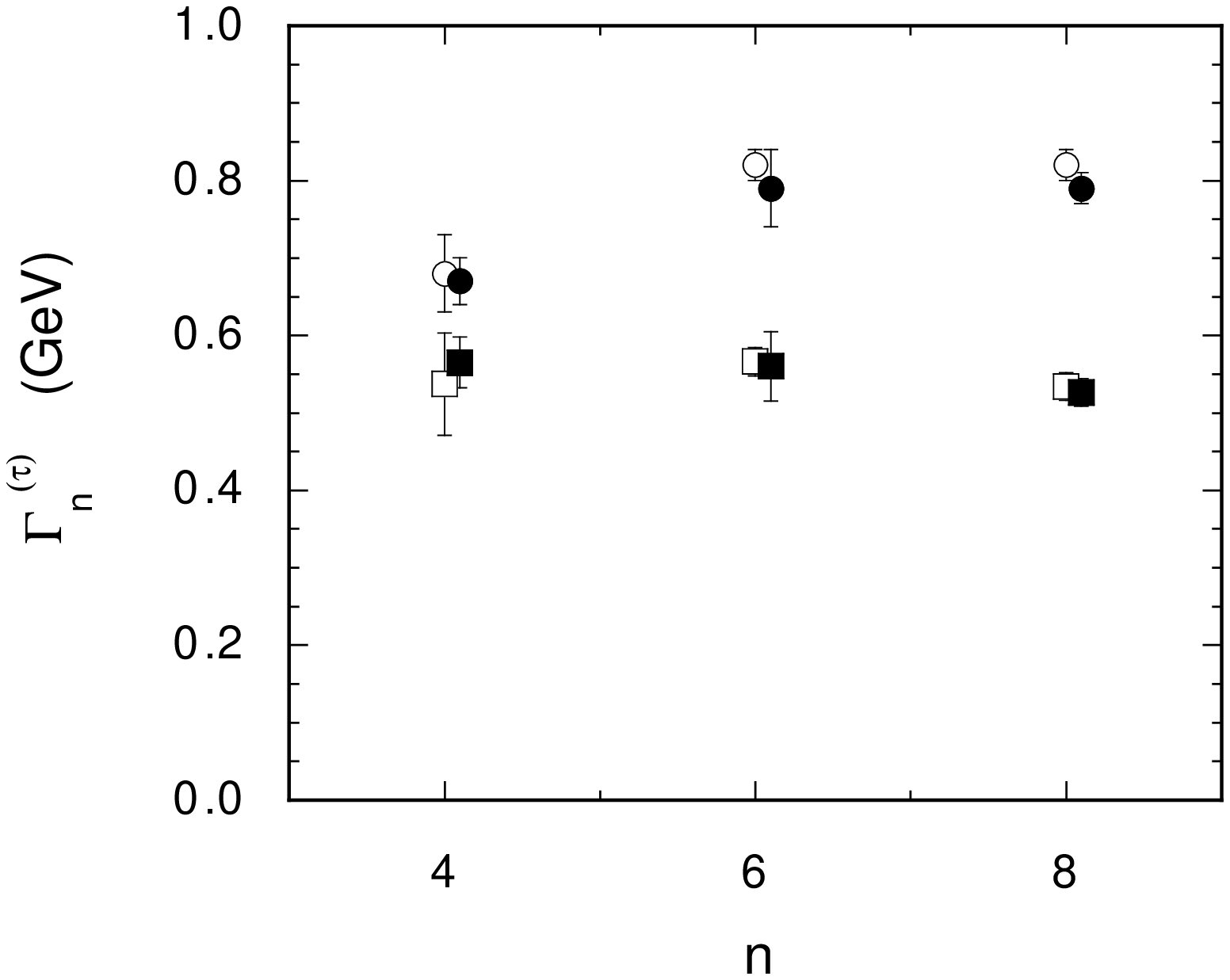}}

\rightline{} \vspace{1cm}

\small{{\bf Figure 13.} The mass scale $\Gamma_n^{(\tau)}$ (see Eq.
(\ref{eq:mass_ht})) of the twist-4 and twist-6 terms resulting from our
final analysis of the transverse pseudo-data (see Tables 6 and 7). Open and 
full markers correspond to the proton and deuteron case, while dots and
squares are our results for the twist-4 and twist-6, respectively (see text).}

\end{figure}

\newpage

\begin{figure}[htb]

\rightline{} \vspace{2cm}

\centerline{\epsfxsize=14cm \epsfig{file=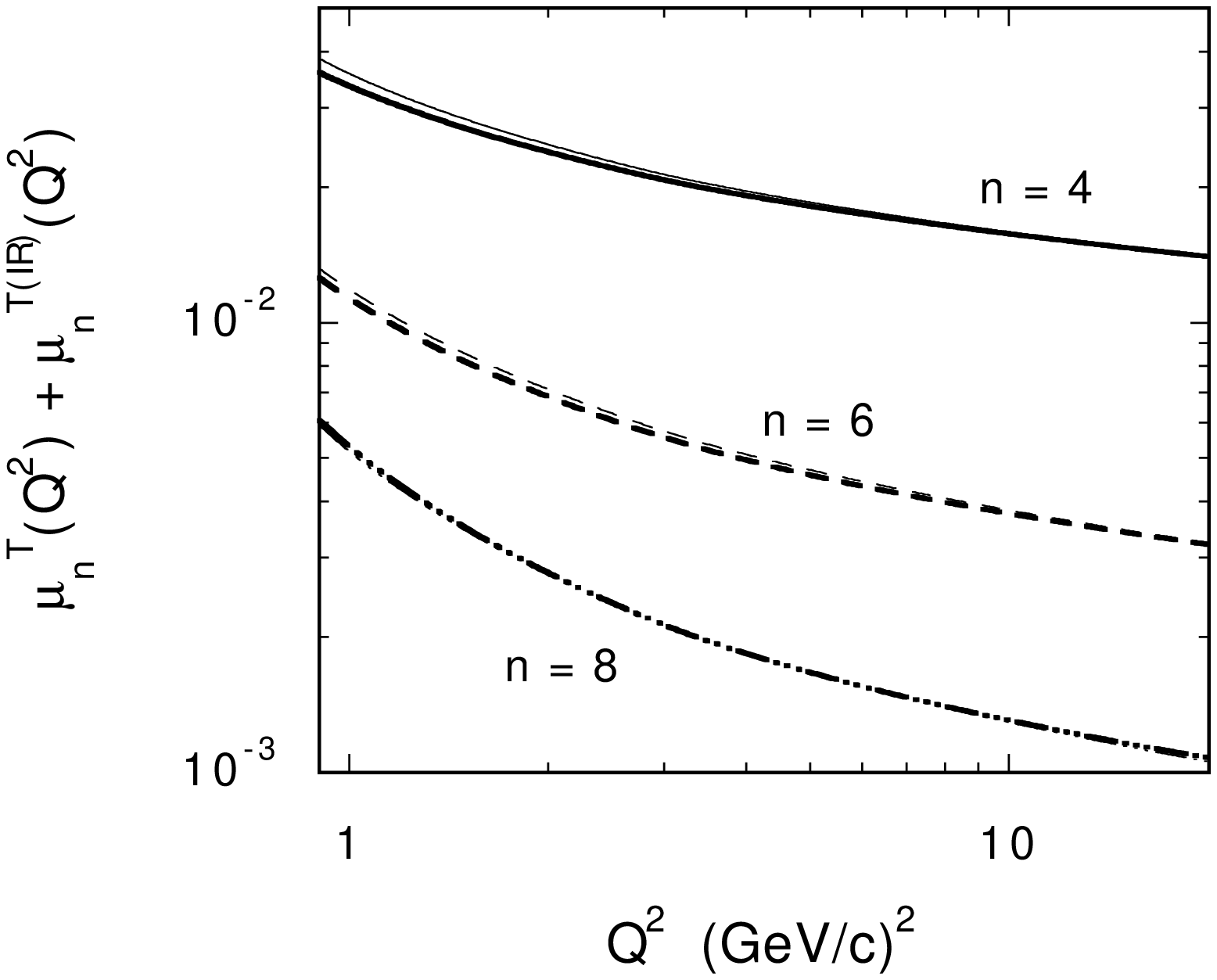}}

\rightline{} \vspace{1cm}

\small{{\bf Figure 14.} The sum of the $NLO$ twist-2 (Eq. (\ref{eq:muT_NLO})) and the $IR$-renormalon (Eq. (\ref{eq:muT_IR})) contributions to the transverse moments $M_n^T(Q^2)$ with $n \geq 4$, as extracted from our final analysis of the transverse pseudo-data (see text), versus $Q^2$. Thin and thick lines are the results obtained at $\alpha_s(M_Z^2) = 0.113$ and $0.118$, respectively, while the solid, dashed and dot-dashed lines correspond to the transverse moments of order $n = 4, 6$ and $8$, respectively.}

\end{figure}

\end{document}